\documentclass[11pt]{article}
\usepackage{amsmath,amssymb,amsthm} 
\usepackage{bbm}
\usepackage{stmaryrd}
\usepackage{paralist}
\usepackage{hyperref}
\usepackage{color}

\setdefaultenum{(i)}{}{}{}
\usepackage[margin=1in]{geometry} 
\usepackage[pdftex]{graphicx}
\usepackage{subfigure}
\theoremstyle{plain}
\newtheorem{mythm}{Theorem} \numberwithin{mythm}{section}
\newtheorem{myprop}[mythm]{Proposition}
\newtheorem{mylemma}[mythm]{Lemma}
\newtheorem{mydef}[mythm]{Definition}
\newtheorem{myrek}[mythm]{Remark}

\def\be{\begin{eqnarray}}
\def\ee{\end{eqnarray}}
\def\b*{\begin{eqnarray*}}
\def\e*{\end{eqnarray*}}

\DeclareMathAlphabet\scr{U}{scr}{m}{n}
\SetMathAlphabet\scr{bold}{U}{scr}{b}{n}
  \DeclareFontFamily{U}{scr}{\skewchar\font'177}%
  \DeclareFontShape{U}{scr}{m}{n}{<-6>rsfs5<6-8>rsfs7<8->rsfs10}{}%
  \DeclareFontShape{U}{scr}{b}{n}{<-6>rsfs5<6-8>rsfs7<8->rsfs10}{}%

\numberwithin{equation}{section}

\renewenvironment{thebibliography}[1]{%
\begin{oldthebibliography}{#1}%
\setlength{\baselineskip}{.8em}
\linespread{.8}
\small
\setlength{\parskip}{0ex}%
\setlength{\itemsep}{.1em}%
}%
{%
\end{oldthebibliography}%
}

\DeclareMathAlphabet\scr{U}{scr}{m}{n}
\SetMathAlphabet\scr{bold}{U}{scr}{b}{n}
  \DeclareFontFamily{U}{scr}{\skewchar\font'177}%
  \DeclareFontShape{U}{scr}{m}{n}{<-6>rsfs5<6-8>rsfs7<8->rsfs10}{}%
  \DeclareFontShape{U}{scr}{b}{n}{<-6>rsfs5<6-8>rsfs7<8->rsfs10}{}%

\def\1{{\bf 1}}
\def\eps{\varepsilon}

\numberwithin{equation}{section}

\def \E{\mathbb{E}}

\def \N{\mathbb{N}}
\def \P{\mathbb{P}}

\def \R{\mathbb{R}}

\def \V{\mathbb{V}}

\def\Ac{{\cal A}}

\def\Cc{{\cal C}}

\def\Fc{{\cal F}}

\def\Ic{{\cal I}}

\def\Nc{{\cal N}}
\def\Oc{{\cal O}}

\def\Rc{{\cal R}}

\def\no{\noindent}

\def\reff#1{{\rm(\ref{#1})}}

\def\1{{\bf 1}}
\def \qed{\hbox{}\hfill {$\square$}}

\def \proof{{\noindent \bf Proof. }}
\def \ep{\hbox{ }\hfill$\Box$}

\def\ep{\epsilon}
\def\eps{\epsilon}

\def\b*{\begin{eqnarray*}}
\def\e*{\end{eqnarray*}}

\begin{document}

\title{\vspace{-1.5cm}Asymptotics for Fixed Transaction Costs\footnote{The authors thank Bruno Bouchard, Jan Kallsen, Ludovic Moreau, Mathieu Rosenbaum, and Peter Tankov for fruitful discussions, and Martin Forde for pertinent remarks on an earlier version. Moreover, they are grateful to two anonymous referees for numerous constructive comments.}}

\author{
Albert Altarovici\thanks{ETH Z\"urich, Departement f\"ur Mathematik, R\"amistrasse 101, CH-8092, Z\"urich, Switzerland, email \texttt{aalbert@math.ethz.ch}.}
\and
Johannes Muhle-Karbe\thanks{ETH Z\"urich, Departement f\"ur Mathematik, R\"amistrasse 101, CH-8092, Z\"urich, Switzerland, and Swiss Finance Institute, email \texttt{johannes.muhle-karbe@math.ethz.ch}. Partially supported by the National Centre of Competence in Research ``Financial Valuation and Risk Management'' (NCCR FINRISK), Project D1 (Mathematical Methods in Financial Risk Management), of the Swiss National Science Foundation (SNF).}
\and H.\ Mete Soner\thanks{ETH Z\"urich, Departement f\"ur Mathematik, R\"amistrasse 101, CH-8092, Z\"urich, Switzerland, and Swiss Finance Institute, email \texttt{hmsoner@ethz.ch}. Partially supported by the
European Research Council under the grant 228053-FiRM and by the ETH Foundation.} 
}

\date{\today}

\maketitle

\begin{abstract}
An investor with constant relative risk aversion trades a safe and several risky assets with constant investment opportunities. For a small \emph{fixed} transaction cost, levied on each trade regardless of its size, we explicitly determine the leading-order corrections to the frictionless value function and optimal policy. 
\end{abstract}




\section{Introduction}

Market frictions play a key role in portfolio choice, ``drastically reducing the frequency and volume of trade'' \cite{constantinides.86}. These imperfections manifest themselves in various forms. Trading costs \emph{proportional} to the traded volume affect all investors in the form of bid-ask spreads. In addition, \emph{fixed} costs, levied on each trade regardless of its size, also play a key role for small investors.

Proportional transaction costs have received most of the attention in the literature. On the one hand, this is due to their central importance for investors of all sizes. On the other hand, this stems from their relative analytical tractability: by their very definition, proportional costs are ``scale invariant'', in that their effect scales with the number of shares traded. With constant relative or absolute risk aversion and a constant investment opportunity set, this leads to a no-trade region of constant width around the frictionless target position \cite{magill.constantinides.76,constantinides.86,davis.norman.90, dumas.luciano.91,shreve.soner.94}. Investors remain inactive while their holdings lie inside this region, and engage in the minimal amount of trading to return to its boundaries once these are breached. The trading boundaries can be determined numerically by solving a free boundary problem \cite{davis.norman.90}. In the limit for small costs, the no-trade region and the corresponding utility loss can be determined explicitly at the leading order, cf.\ Shreve and Soner \cite{shreve.soner.94}, Whalley and Wilmott \cite{whalley.wilmott.97}, Jane{\v{c}}ek and Shreve~\cite{janecek.shreve.04}, as well as many more recent studies \cite{bichuch.11,gerhold.al.11,soner.touzi.12,possamai.al.12,bichuch.shreve.13}.
Extensions to more general preferences and stochastic opportunity sets have been studied numerically by Balduzzi, Lynch, and Tan \cite{lynch.balduzzi.00,balduzzi.lynch.99,lynch.tan.11}. Corresponding formal asymptotics have been determined by Goodman and Ostrov \cite{goodman.ostrov.10}, Martin \cite{martin.12}, Kallsen and Muhle-Karbe \cite{kallsen.muhlekarbe.12,kallsen.muhlekarbe.13} as well as Soner and Touzi \cite{soner.touzi.12}. The last study,  \cite{soner.touzi.12}, also contains a rigorous convergence proof for general utilities, which is extended to several risky assets by Possama\"i, Soner, and Touzi \cite{possamai.al.12}.

Proportional costs lead to infinitely many small transactions. In contrast, fixed costs only allow for a finite number of trades over finite time intervals. However, the optimal policy again corresponds to a no-trade region. In this setting, trades of all sizes are penalized equally, therefore rebalancing takes place by a bulk trade to the optimal frictionless target inside the no-trade region \cite{eastham.hastings.88}. These ``simple'' policies involving only finitely many trades are appealing from a practical point of view. However, fixed costs destroy the favorable scaling properties that usually allow to reduce the dimensionality of the problem for utilities with constant relative or absolute risk aversion. In particular, the boundaries of the no-trade region are no longer constant, even in the simplest settings with constant investment opportunities as well as constant absolute or relative risk aversion. Accordingly, the literature analyzing the impact of fixed trading costs is much more limited than for proportional costs: on the one hand, there are a number of numerical studies \cite{schroder.95,liu.04}, which iteratively solve the dynamic programming equations. On the other hand, Korn~\cite{korn.98} as well as Lo, Mamaysky, and Wang \cite{lo.al.04} have obtained formal asymptotic results for investors with constant \emph{absolute} risk aversion. For small costs, these authors find that constant trading boundaries are optimal at the leading order. Thus, these models are tractable but do not allow us to study how the impact of fixed trading costs depends on the size of the investor under consideration. The same applies to the ``quasi-fixed'' costs proposed by Morton and Pliska~\cite{morton.pliska.95}, and analyzed in the small-cost limit by Atkinson and Wilmott \cite{atkinson.wilmott.95}. In their model, each trade -- regardless of its size -- incurs a cost proportional to the investors' current wealth, leading to a scale-invariant model where investors of all sizes are affected by the ``quasi-fixed'' costs to the same extent. Similarly, the asymptotically efficient discretization rules developed by Fukasawa \cite{fukasawa.11,fukasawa.12} as well as Rosenbaum and Tankov \cite{rosenbaum.tankov.12} also do not take into account that the effect of fixed trading costs should depend on the ``size'' of the investor under consideration.\footnote{Indeed, these schemes asymptotically correspond to constant absolute risk version, cf.\ \cite{fukasawa.12} for more details.} 

The present study helps to overcome these limitations by providing rigorous asymptotic expansions for investors with constant \emph{relative} risk aversion.\footnote{For our \emph{formal} derivations, we consider general utilities like in recent independent work of Alcala and Fahim \cite{alcala.fahim.13}.}  In the standard infinite-horizon consumption model with constant investment opportunities, we obtain explicit formulas for the leading-order welfare effect of small fixed costs and a corresponding almost-optimal trading policy. These shed new light on the differences and similarities compared to proportional transaction costs. 

A universal theme is that, as for proportional transaction costs \cite{janecek.shreve.04,martin.12,kallsen.muhlekarbe.12,kallsen.muhlekarbe.13}, the crucial statistic of the optimal frictionless policy turns out to be its ``portfolio gamma'', which trades off the local variabilities of the strategy and the market (cf.\ \eqref{eq:gamma}). The latter is also crucial in the asymptotic analysis of finely discretized trading strategies \cite{zhang.99,bertsimas.al.00,hayashi.mykland.05,fukasawa.11,fukasawa.12,rosenbaum.tankov.12}. Therefore, it appears to be an appealingly robust proxy for the sensitivity of trading strategies to small frictions.

A fundamental departure from the corresponding results for proportional transaction costs is that the effect of small fixed costs is inversely proportional to investors' wealth. That is, doubling the fixed cost has the same effect on investors' welfare and trading boundaries as halving their wealth.\footnote{Here, both quantities are measured in relative terms, as is customary for investors with constant relative risk aversion. That is, trading boundaries are parametrized by the fractions of wealth held in the risky asset, and the welfare effect is described by the relative certainty equivalent loss, i.e., the fraction of the initial endowment the investor would be willing to give up to trade without frictions.} This explains and quantifies to what extent fixed costs can be neglected by large institutional entities, yet play a key role for small private investors. For example, for typical market parameters (cf.\ Figure \ref{fig:boundaries}), a fixed transaction cost of \$1 per trade leads to trading boundaries of 45\% and 59\% around the frictionless Merton proportion of 52\% if the investor's wealth is \$5000. If wealth increases to \$100000, however, the trading boundaries narrow to 49\% and 55\%, respectively. Our results also show that, asymptotically for small costs, fixed transaction costs are equivalent -- both in terms of the no-trade region and the corresponding welfare loss -- to a suitable ``equivalent proportional cost''. Since the effect of the fixed costs varies with investors' wealth, this equivalent proportional cost is not constant, but decreases with the investors' wealth level. For example, with typical market parameters (cf.\ Figure \ref{fig:boundaries}) a \$1 fixed cost corresponds to a proportional cost of 2.3\% if the investor's wealth is \$5000, but to only 0.24\% if wealth is \$100000. In a similar spirit, our results are also formally linked to those of Atkinson and Wilmott \cite{atkinson.wilmott.95}: their trading costs, taken to be a constant fraction of the investors' current wealth, formally lead to the same results as substituting a stochastic fixed cost proportional to current wealth into our formulas. 

A second novelty is that our results readily extend to a multivariate setting with several risky assets. This is in contrast to the models with proportional transaction costs, where optimal no-trade regions for several risky assets can only be determined numerically by solving a multidimensional nonlinear free-boundary problem, even in the limit for small costs \cite{possamai.al.12}. With small fixed costs, the optimal no-trade region with several risky assets turns out to be an ellipsoid centered around the frictionless target, whose precise shape is easily determined even in high dimensions by the solution of a matrix-valued algebraic Riccati equation. This is again in line with the quasi-fixed costs studied by Atkinson and Wilmott \cite{atkinson.wilmott.95}, up to rescaling the transaction cost by current wealth. Qualitatively, the shape of our ellipsoid resembles the one for the parallelogram-like regions computed numerically for proportional transaction costs by Muthuraman and Kumar \cite{muthuraman.kumar.06} as well as by Possama\"i, Soner, and Touzi \cite{possamai.al.12}. On a quantitative level, however, we find that the shape of the ellipsoid is much more robust with respect to correlation among the risky assets. 

Finally, the present study provides the first rigorous proofs for asymptotics with small fixed costs, complementing earlier partially heuristic results \cite{korn.98,lo.al.04,alcala.fahim.13}, rigorous analyses of the related problem of optimal discretization \cite{fukasawa.11,fukasawa.12,rosenbaum.tankov.12}, and rigorous asymptotics with proportional costs (see \cite{shreve.soner.94,janecek.shreve.04,bichuch.11,gerhold.al.11,soner.touzi.12,possamai.al.12,bichuch.shreve.13}). As for proportional costs \cite{soner.touzi.12}, our approach is based on the theories of viscosity solutions and homogenization, in particular, the weak-limits technique of Barles and Perthame \cite{barles.perthame.87} as well as Evans \cite{evans.89}. However, substantial new difficulties have to be overcome because i) the value function is not concave, ii) the usual dimensionality reduction techniques fail even in the simplest models, iii) the set of controls is not scale invariant, and iv) the dynamic programming equation involves a non-local operator here. In order not to drown these new features in further technicalities, we leave for future research the extension to more general preferences as well as asset price and cost dynamics as in \cite{soner.touzi.12,kallsen.muhlekarbe.12,kallsen.muhlekarbe.13} for proportional costs, and also the analysis of the joint impact of proportional and fixed costs.\footnote{Cf.\ \cite{korn.98,alcala.fahim.13} for corresponding formal asymptotics.}

The remainder of the article is organized as follows: the model, the main results, and their implications are presented in Section 2. Subsequently, we derive the results in an informal manner. This is done in some detail, to explain the general procedure that is likely to be applicable for a number of related problems. In particular, we explain how to come up with the scaling in powers of $\lambda^{1/4}$ by heuristic arguments as in \cite{janecek.shreve.04,rogers.04} and discuss how to use homogenization techniques to derive the corrector equations describing the first-order approximations of the exact solution. Section 4 then makes these formal arguments rigorous by providing a convergence proof. Some technical estimates are deferred to Appendix \ref{app:A}. Finally, Appendix \ref{app:B} presents a self-contained proof of the weak dynamic programming principle, in the spirit of Bouchard and Touzi \cite{bt.11}, which in turn leads to the viscosity solution property of the value function for the problem at hand. 

Throughout, $x^\top$ denotes the transpose of a vector or matrix $x$, $\1_d:=(1,\ldots,1)^\top \in \mathbb{R}^d$, and we write $I_d$ for the identity matrix on $\mathbb{R}^d$. For a vector $x \in \mathbb{R}^d$, $\mathrm{diag}[x]$ represents the diagonal matrix with diagonal elements $x^1,\ldots,x^d$.

\section{Model and Main Results}

\subsection{Market, Trading Strategies, and Wealth Dynamics}

Consider a financial market consisting of a safe asset earning a constant interest rate $r>0$, and $d$ risky assets with expected excess returns $\mu^i-r>0$ and invertible infinitesimal covariance matrix $\sigma \sigma^\top$:
$$dS^0_t=S^0_t r dt, \quad dS_t=S_t \mu dt +S_t \sigma dW_t,$$
for a $d$-dimensional standard Brownian motion $(W_t)_{t \geq 0}$ defined on a filtered probability space $(\Omega,\scr{F},(\scr{F}_t)_{t \geq 0},P)$. Each trade incurs a \emph{fixed transaction cost} $\lambda>0$, regardless of its size or the number of assets involved. As a result, portfolios can only be rebalanced finitely many times, and trading strategies can be described by pairs $(\tau,m)$, where the trading times $\tau=(\tau_1,\tau_2,\ldots)$ are a sequence of stopping times increasing towards infinity, and the $\scr{F}_{\tau_i}$-measurable, $\mathbb{R}^{d}$-valued random variables collected in $m=(m_1,m_2,\ldots)$ describe the transfers at each trading time. More specifically, $m^j_i$ represents the monetary amount transferred from the safe to the $j$-th risky asset at time $\tau_i$. Each trade is assumed to be self-financing, and the fixed costs are deducted from the safe asset account. Thus, the safe and risky positions evolve as 
$$ (x,y)=(x,y^1,\ldots,y^d) \mapsto \left(x-\sum_{j=1}^d m^j_i -\lambda, y^1+m^1_i, \ldots, y^d+m^d_i\right)$$
for each trade $m_i$ at time $\tau_i$. The investor also consumes from the safe account at some rate $(c_t)_{t \geq 0}$. Hence, starting from an initial position $(X_{0-}, Y_{0-}) = (x,y)\in \mathbb{R}\times \mathbb{R}^d$, the wealth dynamics corresponding to a \emph{consumption-investment strategy} $\nu=(c,\tau,m)$ are given by 
\begin{align*}
X_t & = x+ \int_0^t(rX_s - c_s)ds - \sum_{k=1}^{\infty}\left(\lambda + \sum_{j=1}^d m^j_k\right) 1_{\{\tau_k \leq t\}},\\
Y^i_t & =  y^i+ \int_0^tY^i_s\frac{dS^i_s}{S^i_s} +\sum_{k=1}^{\infty}m^i_k 1_{\{\tau_k \leq t\}}.
\end{align*}
We write $(X,Y)^{\nu, x,y}$ for the solution of the above equation. The \emph{solvency region}
$$K_{\lambda} := \left\{(x,y)\in \mathbb{R}^{d+1}:\  \max\left\{\ x+y\cdot \1_d-\lambda\ ,\ \min_{i=1,\ldots, d}\{x,y^i\}\ \right\}\geq 0\right\}.$$
is the set of positions with nonnegative liquidation value. A strategy $\nu = (c, \tau, m)$ starting from the initial position $(x,y)$ is called \emph{admissible} if it remains solvent at all times: $(X_t,Y_t)^{\nu, x,y}\in K_{\lambda}$, for all $t\geq 0$, $P$-a.s. The set of all admissible strategies is denoted by $\Theta^{\lambda}(x,y).$

\subsection{Preferences}

In the above market with constant investment opportunities $(r,\mu,\sigma)$ and fixed transaction costs $\lambda$, an investor with \emph{constant relative risk aversion} $\gamma>0$, i.e., with a utility function $U_{\gamma}:(0,\infty)\to \mathbb{R}$ of either logarithmic or power type,
$$
U_{\gamma}(c) = \begin{cases}
c^{1-\gamma}/(1-\gamma), & 0 <\gamma\neq 1, \\
\log{c}, & \gamma = 1,
\end{cases}
$$
and \emph{impatience rate} $\beta>0$ trades to maximize the expected utility from consumption over an infinite horizon, starting from an initial endowment of $X_{0-} = x$ in the safe and $Y_{0-}=y$ in the risky assets, respectively:\footnote{By convention, the value of the integral is set to minus infinity if its negative part is infinite.}
\begin{equation}\label{eq:vfeps}
v^{\lambda}(x,y) = \sup_{(c,\tau,m)\in \Theta^{\lambda}(x,y)}\mathbb{E}\left[\int_0^{\infty}e^{-\beta t}U_{\gamma}(c_t)dt\right].
\end{equation}

\begin{mythm}\label{thm:visc}
The value function $v^{\lambda}$ of the problem with fixed costs $\lambda>0$ is a (possibly) discontinuous viscosity solution of the Dynamic Programming Equation \eqref{eq:dpe} in the domain 
$$
\Oc_\lambda= \{ (x,y) \in K_\lambda\ :\
x+ y \cdot \1_d > 2 \lambda\}.
$$
\end{mythm}

For our asymptotic results, it suffices to obtain this result for $\Oc_\lambda$ rather than the full solvency region $K_\lambda$. This is because any fixed initial allocation $(x,y)\in \R^{d+1}$ with $x+y\cdot 1_d > 0$ will satisfy $(x,y)\in \Oc_\lambda$ for sufficiently small $\lambda$.

For the definition of a discontinuous viscosity solution, we refer the reader to \cite{crandall.al.92,fleming.soner.06,kabanov.safarian.09,oksendal.sulem.02}. {\O}ksendal and Sulem \cite{oksendal.sulem.02} study existence and uniqueness for one risky asset and power utility with risk aversion $\gamma \in (0,1)$ under the additional assumption $\beta > (1-\gamma)\mu$, a sufficient condition for the finiteness of the frictionless value function. 
The proof of the Theorem \ref{thm:visc} is given in Appendix~\ref{ap:dpp} by establishing a weak dynamic programming principle in the spirit of Bouchard and Touzi \cite{bt.11}. We believe that, in analogy to corresponding results for proportional costs~\cite{shreve.soner.94}, 
the above theorem as well as a comparison result hold in the entire solvency region for
all utility functions whenever the transaction cost value function $v^{\lambda}$ is finite. 
However, this extension is not needed here. 

\subsection{Main Results}\label{sec:main}

Let us first collect the necessary inputs from the frictionless version of the problem (cf., e.g., \cite{fleming.soner.06}): denote by
$$\pi_m=(\sigma \sigma^{\top})^{-1} (\mu-r\1_d)/\gamma$$ 
the optimal frictionless target weights, i.e., the Merton proportions, in the risky assets. Write
$$c_m(\gamma)=\frac{1}{\gamma}\beta+\left(1-\frac{1}{\gamma}\right)\left(r+\frac{(\mu-r\1_d)^{\top}(\sigma \sigma^{\top})^{-1} (\mu-r\1_d)}{2\gamma}\right)$$
for the frictionless optimal consumption rate and let
\begin{equation}\label{eq:valfun0}
v(z)=\begin{cases} \frac{z^{1-\gamma}}{1-\gamma} c_m^{-\gamma}, &\gamma \neq 1,\\
\frac{1}{\beta} \log(\beta z) +\frac{1}{\beta^{2}} \left(r+\frac{(\mu-r\1_d)^{\top}(\sigma \sigma^{\top})^{-1} (\mu-r\1_d)}{2}-\beta)\right), &\gamma=1,
\end{cases}
\end{equation}
be the value function for the frictionless counterpart of \eqref{eq:vfeps} with initial wealth $z=x+y \cdot \1_d$. The latter is finite provided that $c_m>0$, which we assume throughout. Moreover, we also suppose that the matrix
\begin{equation}\label{eq:posdef}
\alpha=(I_d-\pi_m \1_d^\top)\mathrm{diag}[\pi_m]\sigma
\end{equation}
is invertible. 

\begin{myrek}
\rm{
Assuming \eqref{eq:posdef} to be invertible ensures that the asymptotically optimal no-trade region in Theorem \ref{thm:strategy} below is nondegenerate. This is tantamount to a non-trivial investment in each of the $d+1$ assets.
}
\end{myrek}

Our main results are the leading-order corrections for small fixed transaction costs $\lambda$; their interpretation as well as connections to the literature are discussed in Section \ref{sec:interpretations} below.

\begin{mythm}[Expansion of the Value Function]\label{t.main}
For all solvent initial endowments $(x,y)\in \mathbb{R}^{d+1}$ with $z=x+y\cdot \1_d > 0$, we have
\[v^{\lambda}(x,y) = v(z)-\lambda^{1/2}u(z) + o(\lambda^{1/2}),\]
that is,
$$
u^{\lambda}(x,y) := \frac{v(z) - v^{\lambda}(x,y)}{\lambda^{1/2}}\to u(x+y),
$$
locally uniformly as $\lambda\to 0$. Here, 
$$u(z)=u_0 z^{1/2-\gamma},$$
for a constant $u_0>0$ determined by the \emph{corrector equations} from Definition \ref{def:correct}. For a single risky asset ($d=1$):
$$u_0=\sigma^2 \left(\frac{\gamma}{3} \pi_m^2(1-\pi_m)^2\right)^{1/2} \frac{c_m(\gamma)^{-\gamma}}{c_m(2\gamma)};$$
see Section \ref{sec:multi} for the multivariate case. This determines the leading-order \emph{relative certainty equivalent loss}, i.e., the fraction of her initial endowment the investor would give up to trade the risky asset without transaction costs, as follows:
\begin{equation}\label{eq:rce}
v^\lambda(x,y) = v\left(z\left(1-u_0 c_m(\gamma)^{\gamma} \frac{\lambda^{1/2}}{z^{1/2}}\right)\right)+o(\lambda^{1/2}).
\end{equation}
\end{mythm}

The leading-order optimal performance from Theorem \ref{t.main} is achieved by the following ``almost optimal policy'':

\begin{mythm}[Almost Optimal Policy]\label{thm:strategy}
Fix a solvent initial portfolio allocation. Define the no-trade region
$$\mathrm{NT}^\lambda=\left\{(x,y) \in \mathbb{R}^{d+1}:  \frac{y}{x+y \cdot \1_d}  \in \pi_m +  \frac{\lambda^{1/4}}{(x+y \cdot \1_d)^{1/4}} \scr{J}\right\},$$
for the ellipsoid $\scr{J}=\{\rho \in \mathbb{R}^d: \rho^\top M\rho <1 \}$ from Section \ref{sec:multi}. Consider the strategy which consumes at the frictionless Merton rate, does not trade while the current position lies in the above no-trade region, and jumps to the frictionless Merton proportion once its boundaries are breached. Then, for any $\delta>0$, the utility obtained from following this strategy until wealth falls to level $\delta$, and then switching to a leading-order optimal strategy for \eqref{eq:vfeps}, is optimal at the leading order $\lambda^{1/2}$ (cf.\ Section \ref{s.almost} for more details).

For a single risky asset, the above no-trade region simplifies to the following interval around the frictionless Merton proportion:
\begin{equation}\label{eq:almost}
\mathrm{NT}^\lambda=\left\{(x,y) \in \mathbb{R}^2:  \left|\frac{y}{x+y} -\pi_m \right| \leq  \left(\frac{12}{\gamma}\pi_m^2(1-\pi_m)^2 \frac{\lambda}{x+y}\right)^{1/4}\right\}.
\end{equation}
\end{mythm}

\begin{myrek}\label{r.threshold}\rm{
Unlike for proportional transaction costs, trading only after leaving the above asymptotic no-trade region is not admissible for any given fixed cost $\lambda>0$. This is because wealth can fall below the level $\lambda$ needed to perform a final liquidating trade. Hence, the above region is only ``locally'' optimal, in that one needs to switch to the unknown optimal policy after wealth falls below a given threshold.}
\end{myrek}

\subsection{Interpretations and Implications}\label{sec:interpretations}

In this section, we discuss a number of interpretations and implications of our main results. We first focus on the simplest case of one safe and one risky asset, before turning to several correlated securities.

\subsubsection*{Small Frictions and Portfolio Gammas}

The transactions of the optimal policies for proportional and fixed costs are radically different. For proportional costs, there is an infinite number of small trades of ``local-time type'', whereas fixed costs lead to finitely many bulk trades. Nevertheless, the respective no-trade regions -- that indicate when trading is initiated -- turn out to be determined by exactly the statistics summarizing the market and preference parameters.

Indeed, just as for proportional transaction costs \cite{janecek.shreve.04}, the width of the leading-order optimal no-trade region in \eqref{eq:almost} is determined by a power of $\pi_m^2(1-\pi_m)^2$ rescaled by the investor's risk tolerance $1/\gamma$. This term quantifies the sensitivity of the current risky weight with respect to changes in the price of the risky asset, cf.\ \cite[Remark 4]{janecek.shreve.04}. Compared to the corresponding formula for proportional transaction costs in \cite{janecek.shreve.04}, it enters through its quartic rather than cubic root, and is multiplied by a different constant. Nevertheless, most qualitative features remain the same: the leading-order no-trade region vanishes if a full safe or risky investment is optimal in the absence of frictions ($\pi_m=0$ or $\pi_m=1$, respectively) and the effect on optimal strategies increases significantly in the presence of leverage ($\pi_m>1$, compare \cite{gerhold.al.11}). 

As in \cite{martin.12,kallsen.muhlekarbe.12,kallsen.muhlekarbe.13} for proportional costs, the no-trade region can also be interpreted in terms of the activities of the frictionless optimizer and the market as follows. Let $\varphi_m(t)=\pi_m Z_t/S_t$ be the frictionless optimal strategy for current wealth $Z_t$, expressed in terms of the number of shares held in the risky asset. Then, the frictionless wealth dynamics $dZ_t=Z_t \pi_m dS_t/S_t -c_t dt$ and It\^o's formula yield
$$\frac{d\langle\varphi_m\rangle_t}{dt} =\frac{\pi_m^2(1-\pi_m)^2 \sigma^2 Z_t^2}{S_t^2}.$$
As a result, the maximal deviations \eqref{eq:almost} from the frictionless target can be rewritten in numbers of risky shares as 
$$\pm\left(\frac{12}{\gamma} \frac{d\langle \varphi_m \rangle_t}{d\langle S \rangle_t}\frac{\lambda}{Z_t}\right)^{1/4}.$$
Our formal results from Section \ref{sec:sol1} suggest that an analogous result remains valid also for more general preferences. Then, the frictionless target $\varphi_m(t)=\theta(Z_t)/S_t$ (cf.\ Section \ref{sec:nofric}) is no longer constant, and It\^o's formula yields
$$\frac{d\langle \varphi_m \rangle_t}{dt}=\frac{\sigma^2 \theta^2(Z_t)(1-\theta_z^2(Z_t))^2}{S_t^2},$$
so that the maximal deviations  \eqref{eq:dev}  from the frictionless target $\varphi_m(t)$ can be written as
\begin{equation}\label{eq:gamma}
\pm\left(\frac{12}{-v_{zz}(z)/v_z(z)} \frac{d\langle \varphi_m \rangle_t}{d\langle S \rangle_t}\lambda\right)^{1/4},
\end{equation}
in terms of numbers of risky shares. Up to changing the power and the constant, this is the same formula as for proportional transaction costs \cite{kallsen.muhlekarbe.12,martin.12,kallsen.muhlekarbe.13}: the width of the no trade region is determined by the transaction cost, times the (squared) \emph{portfolio gamma} $d\langle \varphi_m \rangle_t/d\langle S \rangle_t$, times the risk-tolerance of the indirect utility function of the frictionless problem. The portfolio gamma also is the key driver in the analysis of finely discretized trading strategies \cite{zhang.99,bertsimas.al.00,hayashi.mykland.05,fukasawa.11,fukasawa.12,rosenbaum.tankov.12}. Hence, it appears to be an appealingly robust measure for the sensitivity of trading strategies to small frictions.

\subsubsection*{Wealth Dependence and Equivalent Proportional Costs}

\begin{figure}
\begin{center}
\includegraphics[height=1.9in]{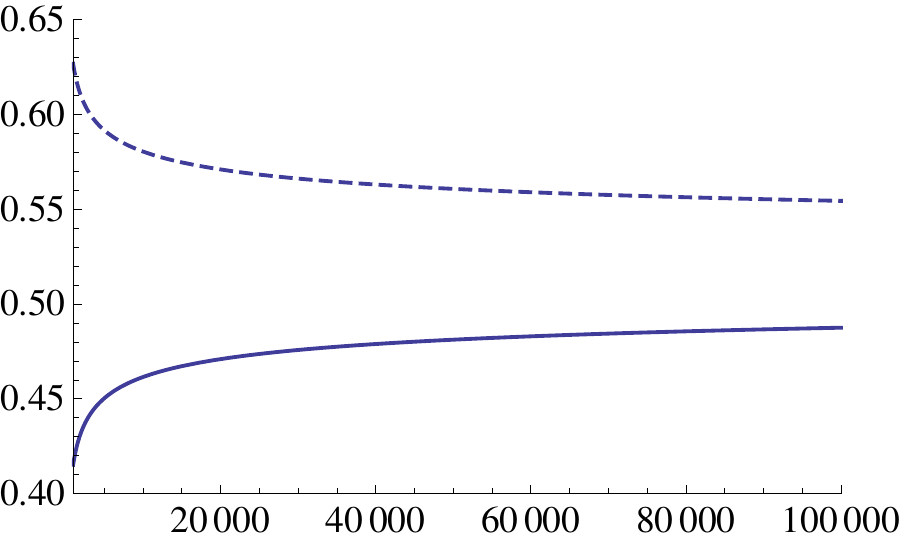}
\includegraphics[height=1.9in]{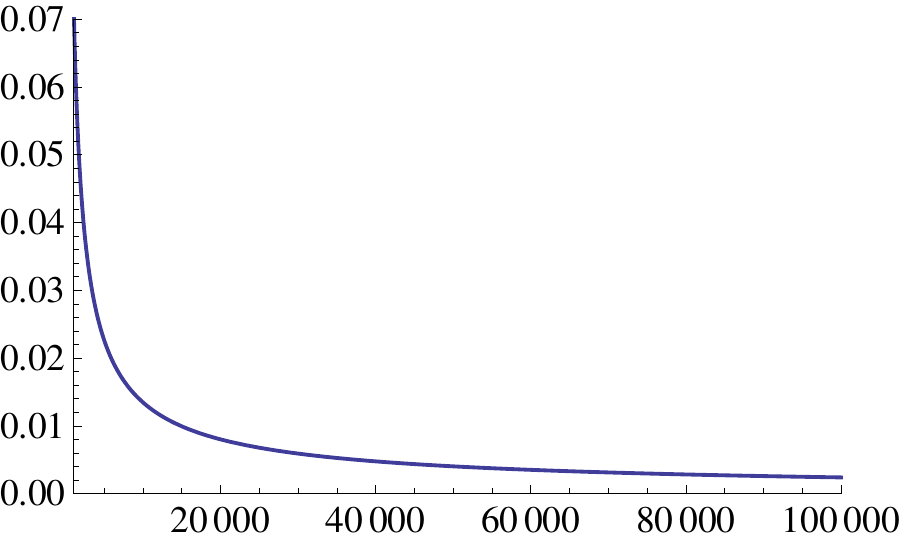}
\caption{\label{fig:boundaries}
Trading boundaries (left panel) and equivalent proportional costs (right panel) for a \$1 fixed transaction cost, as functions of the investor's wealth. Parameters are $\gamma=6$, $\mu-r=8\%$, and $\sigma=16\%$, so that the Merton proportion is $\pi_m=52\%$}
\end{center}
\end{figure}

A fundamental departure from the corresponding results for proportional transaction costs is that the impact of fixed costs depends on investors' wealth. Indeed, the fixed cost $\lambda$ is normalized by the investors' current wealth, both in the asymptotically optimal trading boundaries \eqref{eq:almost} and in the leading-order relative welfare loss \eqref{eq:rce}, see Figure \ref{fig:boundaries} for an illustration. This makes precise to what extent fixed costs can indeed be neglected for large institutional traders, but play a key role for small private investors: ceteris paribus, doubling the investors' wealth reduces the impact of fixed trading costs in exactly the same way as halving the costs themselves. As a result, a constant fixed cost leads to a no-trade region that fluctuates with the investors' wealth. In contrast, for proportional transaction costs, this only happens if these evolve stochastically. The formal results of Kallsen and Muhle-Karbe \cite{kallsen.muhlekarbe.13} shed more light on this connection. It turns out that a constant fixed cost $\lambda$ is equivalent -- both in terms of the associated no-trade region \emph{and} the corresponding welfare loss -- to a random and time-varying proportional cost given by
$$\lambda_t^{\mathrm{equiv}}=\left(\frac{1024 \gamma}{3 \pi_m^2 (1-\pi_m)^2}\right)^{1/4} \left(\frac{\lambda}{Z_t}\right)^{3/4},$$
for current total wealth $Z_t$.\footnote{To see this, formally let the time horizon tend to infinity in \cite[Sections 4.1 and 4.2]{kallsen.muhlekarbe.13} and insert the explicit formulas for the optimal consumption rate and risky weight. This immediately yields that the leading-order no-trade regions coincide; for the corresponding welfare effects this follows after integrating.} Note that this formula is independent of the impatience parameter $\beta$, and only depends on the market parameters ($\mu,\sigma,r$) through the Merton proportion $\pi_m=(\mu-r)/\gamma\sigma^2$. This relation clearly shows that a fixed cost corresponds to a larger proportional cost if rebalancing trades are small because i) the investors' wealth $Z_t$ is small or ii) the no-trade region is narrow because the frictionless optimal position $\pi_m$ is close to a full safe or risky position ($\pi_m=0$ or $\pi_m=1$). In contrast, for large investors and a frictionless position sufficiently far away from full risky or safe investment, the effect of fixed costs becomes negligible (cf.\ Figure 1 for an illustration). For sufficiently high risk aversion $\gamma$, the equivalent proportional cost is increasing in risk aversion (as higher risk aversion leads to smaller trades), in line with the numerical findings of Liu \cite{liu.04} for exponential utility. Here, however, one can additionally assess the impact of changing wealth over time endogenously, rather than by having to vary the investors' risk aversion. 

Our asymptotic formulas for fixed costs also allow to relate these to the fixed \emph{fraction} of current wealth charged per transaction in the model of Morton and Pliska \cite{morton.pliska.95}. Their ``quasi-fixed'' costs are scale-invariant, in that they lead to constant trading boundaries around the Merton proportion $\pi_m$, whose asymptotics have been derived by Atkinson and Wilmott \cite{atkinson.wilmott.95}.  Formally, these trading boundaries coincide with ours if the ratio of their time-varying trading cost and our fixed fee  is given by the investors' current wealth.

\subsubsection*{Multiple Stocks}

For multiple stocks, Theorem \ref{thm:strategy} shows that it is approximately optimal to keep the portfolio weight in an ellipsoid around the frictionless Merton position $\pi_m$. Whereas nonlinear free-boundary problems have to be solved to determine the optimal no-trade region for proportional costs even if these are small \cite{possamai.al.12}, the asymptotically optimal no-trade ellipsoid with fixed costs is determined by a matrix-valued algebraic Riccati equation, which is readily evaluated numerically even in high dimensions (see Section~\ref{sec:multi} for more details). Qualitatively, this is again in analogy to the asymptotic results of Atkinson and Wilmott \cite{atkinson.wilmott.95} for the Merton and Pliska model \cite{atkinson.wilmott.95} but -- as for a single risky asset -- the trading boundary varies with investors' wealth for the fixed costs considered here.

\begin{figure}
\begin{center}
\includegraphics[width=3.15in]{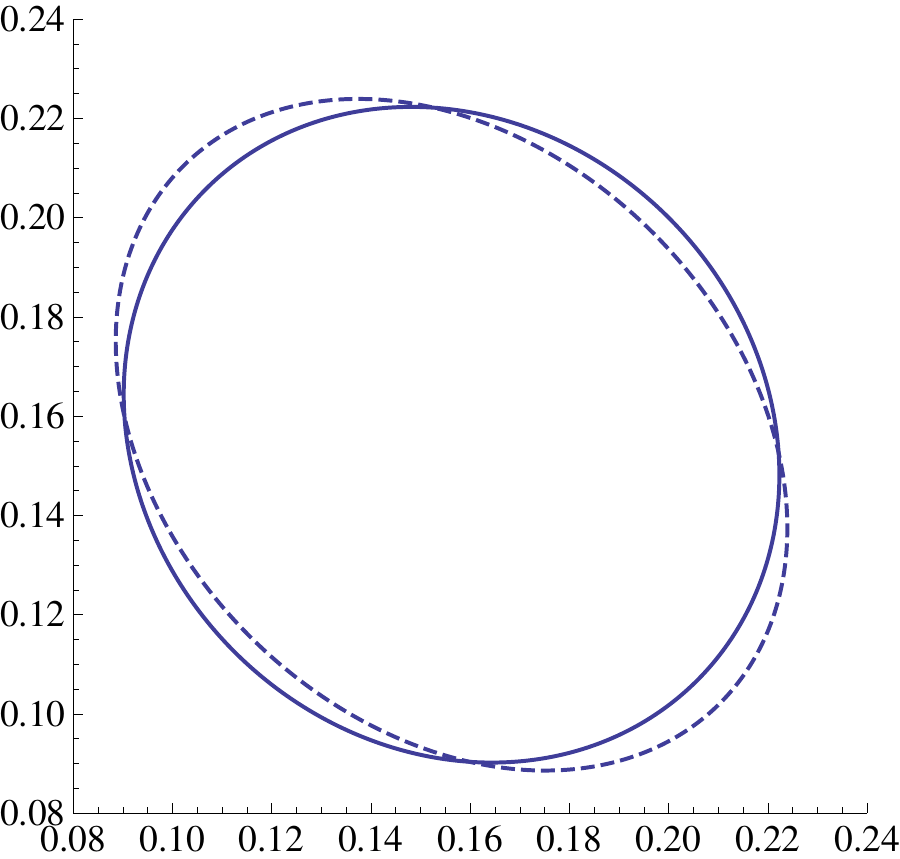}
\caption{\label{fig:corr}No-trade ellipsoid for two identical risky assets with excess returns 5\%, correlation 0 (solid) resp.\ 44\% (dashed) and corresponding volatilities 40\% resp.\ 33\% chosen so that the Merton proportion remains constant at $(5/32,5/32)$. Risk aversion is $\gamma=2$, wealth is $\$50000$, and the fixed cost is \$3.41. }
\end{center}
\end{figure}

\begin{figure}
\begin{center}
\includegraphics[width=3.15in]{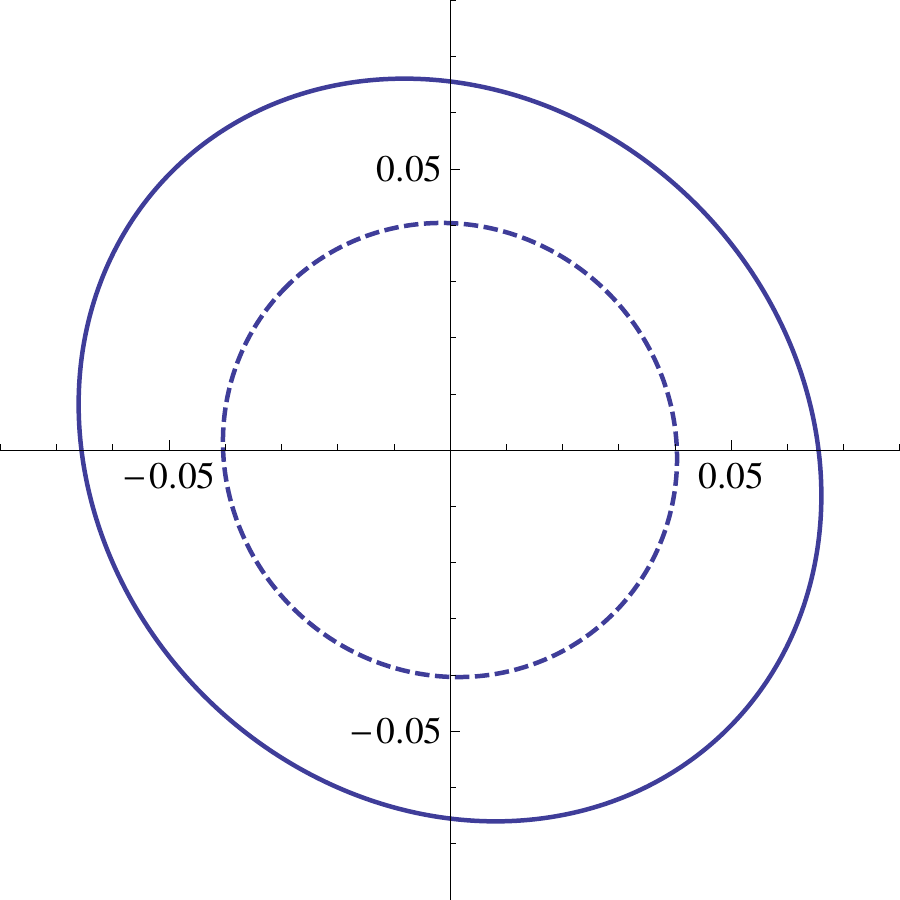}
\caption{\label{fig:ra}Maximal deviations (in percentages of wealth held in the risky assets) from the frictionless Merton proportion for two identical risky assets with excess returns 5\%, correlation 0, volatilities 40\%, and risk aversions 2 (solid) and 6 (dashed). Wealth is $\$50000$, and the fixed cost is \$3.41. }
\end{center}
\end{figure}

To shed some light on the quantitative features of the solution, Figure \ref{fig:corr} depicts the no-trade ellipsoid for two identical  risky assets with varying degrees of correlation.\footnote{To facilitate comparison, we use the same market parameters $\mu,\sigma,r$ and risk aversion $\gamma$ as in Muthuraman and Kumar \cite{muthuraman.kumar.06}. The fixed cost and the current wealth are chosen so that the one dimensional no-trade region for each asset corresponds to the one for their 1\% proportional cost.}  Qualitatively, correlation deforms the shape of the no-trade region similarly as in Muthuraman and Kumar \cite[Figure 6.8]{muthuraman.kumar.06} for proportional costs: in the space of risky asset weights, the no-trade region shrinks in the $(1,1)$ direction but widens in the $(1,-1)$-direction because investors use the positively correlated assets as partial substitutes for one another. 

On a quantitative level, however, the impact of correlation turns out to be considerably less pronounced for fixed costs. This is because whenever \emph{any} trade happens all stocks can be traded with no extra cost, weakening the incentive to use substitutes for hedging. 
Also notice that the no-trade region is not rotationally symmetric even for two identical uncorrelated stocks. This is in contrast to the results for exponential utilities, for which the investor's maximization problem factorizes into a number of independent subproblems \cite{liu.04}. 
Note, however, that as risk aversion rises the optimal no-trade region for uncorrelated identical stocks quickly becomes more and more symmetric, in line with the high risk aversion asymptotics linking power utilities to their exponential counterparts.\footnote{Compare Nutz \cite{nutz.12} for a general frictionless setting, as well as Guasoni and Muhle-Karbe \cite{guasoni.muhlekarbe.12} for a model with proportional transaction costs. A similar result for fixed costs is more difficult to formulate, because the investor's wealth does not factor out of the trading policy in this case.} This is illustrated in Figure \ref{fig:ra}.

\section{Heuristic Derivation of the Solution}

In this section, we explain how to use the homogenization approach to determine the small-cost asymptotics on an informal level. The derivations are similar to the ones for proportional costs~\cite{soner.touzi.12}.

Since this entails few additional difficulties on a \emph{formal} level, we consider general utilities $U$ defined on the positive half-line in this section. For the rigorous convergence proofs in Section 4, we focus on utilities $U_\gamma$ with constant relative risk aversion in order not to drown the  arguments in technicalities.  
\subsection{The Frictionless Problem}\label{sec:nofric}

The starting point for the present asymptotic analysis is the solution of the frictionless version of the problem at hand. Since trades are costless in this setting, the corresponding value function does not depend separately on the positions $x,y$ in the safe and the risky assets, but only on total wealth $z=x+ y \cdot \1_d$. As is well known (cf., e.g., \cite[Chapter X]{fleming.soner.06}), the frictionless value function solves the dynamic programming equation
\begin{equation}\label{eq:dpe0}
0=\widetilde{U}(v_z(z))-\beta v(z)+\scr{L}_0 v(z),
\end{equation}
where
\begin{equation}\label{eq:L0}
\scr{L}_0 v(z)=  v_z(z) z r+ v_z(z) (\mu-r \1_d) \cdot \theta(z)+\frac{1}{2} v_{zz}(z) |\sigma^{\top} \theta(z)|^2,
\end{equation}
and the corresponding optimal consumption rate and optimal risky positions are given by
\begin{equation}\label{eq:c0}
\kappa(z)=(U')^{-1}(v_z(z))
\end{equation}
and
\begin{equation}\label{eq:pi0}
\theta(z):= -\frac{v_z(z)}{v_{zz}(z)} (\sigma \sigma^\top)^{-1} (\mu-r\1_d).
\end{equation}
For power or logarithmic utilities $U_\gamma(z)$ with constant relative risk aversion $-zU_\gamma''(z)/U_\gamma'(z)=\gamma$, this leads to the explicit formulas from Section \ref{sec:main} because the value function is homothetic in this case: $v(z)=z^{1-\gamma}v(1)$ (if $\gamma \neq 1$) resp.\ $v(z)=\frac{1}{\beta}\log(z)+v(1)$ (if $\gamma=1$).

\subsection{The Frictional Dynamic Programing Equation}

For the convenience of the reader, we now recall how to heuristically derive the dynamic programming equation with fixed trading costs. We start from the ansatz that  the value function $v^\lambda(x,y)$ for our infinite horizon problem with constant model parameters should only depend on the positions in each of the assets. Evaluated along the positions $X_t,Y_t$ corresponding to any admissible policy $(c,\tau,m)$, It\^o's formula in turn yields
\begin{equation}\label{eq:vfunc}
\begin{split}
dv^\lambda(X_t,Y_t)=&\left(v^\lambda_x(X_t,Y_t) (rX_t-c_t)+\mu \cdot \mathbf{D}_y v^\lambda(X_t,Y_t) + \frac{1}{2} \mathrm{Tr}(\sigma \sigma^\top \mathbf{D}_{yy} v^\lambda(X_t,Y_t)) \right)dt\\
& + \mathbf{D}_y v^\lambda(X_t,Y_t)^\top \sigma dW_t 
+\sum_{\tau_i \leq t} \left(v^\lambda\left(X_{\tau_i}-m_i \cdot \1_d -\lambda, Y_{\tau_i}+m_i\right)-v^\lambda(X_{\tau_i},Y_{\tau_i})\right),
\end{split}
\end{equation}
where
$$\mathbf{D}^i_y= y^i \frac{\partial}{\partial y_i}, \quad \mathbf{D}^{ij}_{yy}=y^i y^j \frac{\partial^2}{\partial y^i \partial y^j}, \quad i,j=1,\ldots,d.$$
By the martingale optimality principle of stochastic control, the utility
$$\int_0^t e^{-\beta s} U(c_s) ds + e^{-\beta t} v^\lambda(X^{(c,\tau,m)}_t,Y^{(c,\tau,m)}_t)$$
obtained by applying an arbitrary policy $(c,\tau,m)$ until some intermediate time $t$ and then trading optimally should always lead to a supermartingle, and to a martingale if the optimizer is used all along. Between trades -- in the policy's ``no-trade region'' -- this means that the absolutely continuous drift should be nonpositive, and zero for the optimizer. After taking into account \eqref{eq:vfunc}, using integration by parts, and canceling the common factor $e^{-\beta t}$, this leads to
\begin{equation}\label{eq:drift}
0=\sup_{c} \left\{-\beta v^\lambda(x,y)+U(c)-v^\lambda_x(x,y) c +v^\lambda_x(x,y) r x + \mu \cdot \mathbf{D}_y v^\lambda(x,y) + \frac{1}{2}\mathrm{Tr}(\sigma \sigma^\top \mathbf{D}_{yy}) v^\lambda(x,y)\right\}.
\end{equation}
By definition, the value function can only be decreased by admissible bulk trades at any time:
\begin{equation*}\label{eq:bulk}
0 \geq  \sup_{m} \left\{ v^\lambda(x-m \cdot \1_d -\lambda,y+m)-v^\lambda(x,y)\right\},
\end{equation*}
and this inequality should become an equality  for the optimal transaction once the boundaries of the no-trade region are breached. Combining this with \eqref{eq:drift} and switching the sign yields the \emph{dynamic programming equation}:
\begin{equation}\label{eq:dpe}
0= \min\left\{\beta v^\lambda -\widetilde{U}(v^\lambda_x)-\scr{L}v^\lambda, v^\lambda-\mathbf{M}v^\lambda\right\},
\end{equation}
where $\widetilde{U}(y)=\sup_{x>0}(U(x)-xy)$ is the convex dual of the utility function $U$, the differential operator $\scr{L}$ is defined as
$$\scr{L}=r x \frac{\partial}{\partial x}+ \mu \cdot \mathbf{D}_y +\frac{1}{2} \mathrm{Tr}(\sigma\sigma^\top \mathbf{D}_{yy}),$$
and $\mathbf{M}$ denotes the non-local \emph{intervention operator} 
$$\mathbf{M}\psi(x,y)=\sup_{m \in \mathbb{R}^d} \left\{ \psi(x',y'): (x',y')=(x-m \cdot \1_d -\lambda, y+m) \in K_\lambda\right\}.$$

\subsection{Identifying the Correct Scalings}

The next step is to determine heuristically how the optimal no-trade region around the frictionless solution and the corresponding utility loss should scale with a \emph{small} transaction cost $\lambda$. This can be done by adapting the heuristic argument in \cite{janecek.shreve.04,rogers.04}. Indeed, the welfare effect of any trading cost is composed of two parts, namely the direct costs incurred due to actual trades and the displacement loss due to having to deviate from the frictionless optimum. Since the frictionless value function is locally quadratic around its maximum, Taylor's theorem suggests that the displacement effect should be of order $x^2$ for any small cost that only causes a small displacement $x$. Where the various cost structures differ is in the losses due to actual trades. Proportional transaction costs lead to trading of local-time type, which scales with the inverse of the width of the no-trade region \cite[Section 3]{janecek.shreve.04}. This leads to a total welfare loss proportional to
$$C x^2+\lambda /x,$$
for some constant $C>0$. Minimizing this expression leads to a no-trade region with width of order $\lambda^{1/3}$ and a corresponding welfare loss of order $\lambda^{2/3}$. In contrast, trades of all sizes are penalized alike by fixed costs. This leads to a bulk trade to the optimal frictionless position, and therefore a transaction cost of $\lambda$, whenever the boundaries of the no-trade region are reached. On the short time-interval before leaving a narrow no-trade region, any diffusion resembles a Brownian motion at the leading order. Hence, the first exit time can be approximated by the one of a Brownian motion from the interval $[-x,x]$, which scales with $x^2$. After the subsequent jump to the midpoint of the no-trade region, this procedure is repeated, so that the number of trades approximately scales with $1/x^2$. As a result, the total welfare loss due to small fixed costs $\lambda$ is proportional to
$$C x^2 +\lambda/x^2,$$
for  some constant $C>0$. Minimizing this expression in $x$ then leads to an optimal no-trade region of order $\lambda^{1/4}$ and a corresponding welfare loss of order $\lambda^{1/2}$.

\subsection{Derivation of the Corrector Equations}
In view of the above considerations, we expect the leading-order utility loss due to small transaction costs $\lambda$ to be of order $\lambda^{1/2}$, whereas the deviations of the optimal policy from its frictionless counterpart should be of order $\lambda^{1/4}$. This motivates the following ansatz for the asymptotic expansion of the transaction cost value function:
\begin{equation}\label{eq:ansatz}
v^\lambda(x,y)=v(z)-\lambda^{1/2} u(z)-\lambda w(z,\xi) + o(\lambda^{3/4}).
\end{equation}
Here, $v$ is the frictionless value function from Section \ref{sec:nofric}, the functions $u$ and $w$ are to be calculated, and we change variables from the safe and risky positions $x,y$ to total wealth 
$$z:=x+y \cdot \1_d$$ 
and the deviations 
$$\xi:=(y-\theta(x+y))/\lambda^{1/4}$$ 
of the risky positions from their frictionless targets, normalized to be of order $O(1)$ as $\lambda\downarrow 0$. The function $\lambda w$ is included, even though it only contributes at the higher order $\lambda$ itself, because its second derivatives with respect to the $y$-variables are of order $\lambda^{1/2}$. 

To determine $u$ and $w$, insert the postulated expansion \eqref{eq:ansatz} into the Dynamic Programming Equation \eqref{eq:dpe}. This leads to two separate equations in the no-trade and trade region, respectively.

\subsubsection*{No-Trade Region}
To ease notation, we illustrate the calculations for the case of a single risky asset ($d=1$), and merely state the multi-dimensional results at the end.\footnote{The full multi-dimensional derivation can be found in \cite{soner.touzi.12}. In the no-trade region, the calculations are identical.} In the no-trade region, we have to expand the elliptic operator from \eqref{eq:dpe} in powers of $\lambda$. To this end, Taylor expansion, \eqref{eq:c0}, and $\widetilde{U}'=-(U')^{-1}$ yield 
$$\widetilde{U}(v^\lambda_x(x,y))=\widetilde{U}(v_z(z))+\lambda^{1/2} \kappa(z) u_z(z) +o(\lambda^{3/4}).$$
Moreover, also taking into account that $y=\theta(z) +\lambda^{1/4} \xi$, it follows that
\begin{align*}
\beta v^\lambda(x,y) &-\widetilde{U}(v^\lambda_x(x,y))-\scr{L}v^\lambda(x,y)\\
=&\beta v(z) -\widetilde{U}(v_z(z))-\scr{L}_0 v(z)\\
&-\lambda^{1/4}\xi \left(\mu v_z(z)+\sigma^2 \theta(z) v_{zz}(z) \right)\\
&- \lambda^{1/2} \left(\beta u(z) -\scr{L}_0 u(z) + \kappa(z)u_z(z) +\frac{\sigma^2}{2} \xi^2 v_{zz}(z) -\frac{\sigma^2}{2}\theta(z)^2(1-\theta_z(z))^2 w_{\xi\xi}(z,\xi)\right)\\
&+o(\lambda^{1/2}),
\end{align*}
for the differential operator $\scr{L}_0$ from \eqref{eq:L0}. The $O(\lambda^{1/4})$-terms in this expression vanish by definition \eqref{eq:pi0} of the frictionless optimal weight; the same holds for the $O(1)$-terms by the frictionless Dynamic Programming Equation \eqref{eq:dpe0}. Satisfying the elliptic part of equation \eqref{eq:dpe} between bulk trades -- at the leading order $O(\lambda^{1/2})$ -- is therefore tantamount to 
\begin{equation}\label{eq:drift2}
0=\beta u(z) -\scr{L}_0 u(z) + \kappa(z)u_z(z) +\frac{\sigma^2}{2}\xi^2 v_{zz}(z) -\frac{\sigma^2}{2}\theta(z)^2(1-\theta_z(z))^2 w_{\xi\xi}(z,\xi).
\end{equation}

\subsubsection*{Trade Region}
Now, turn to the second part of the frictional Dynamic Programming Equation \eqref{eq:dpe}, which should vanish when a bulk trade becomes optimal outside the no-trade region. Suppose that $(x,y)\in K_{\lambda}$ and $v^{\lambda}(x,y) = \mbox{\bf M}v^{\lambda}(x,y)$. Then, inserting the expansion for $v^{\lambda}$ yields
$$
v(z) - \lambda^{1/2}u(z) - \lambda w(z,\xi) = v(z-\lambda) - \lambda^{1/2}u(z-\lambda) - \lambda\cdot \textstyle{\inf_{\hat{\xi}}w(z-\lambda,\hat{\xi})},
$$
where the infimum is over deviations $\hat \xi$ attainable from the current position $(z,\xi)$ by a single trade. Taylor expansion yields 
$$
0 = \lambda\left(v_z(z) - w(z,\xi) + \textstyle{\inf_{\hat\xi}w(z-\lambda, \hat \xi)}\right) + o(\lambda).
$$
If $w(z,\xi) = w(z-\lambda,\xi) + o(\lambda),$ where $o(\lambda)$ only depends on $z$,\footnote{This will turn out to be consistent with the results of our calculations below; see Sections \ref{sec:sol1} and \ref{sec:multi}.} this simplifies to
$$
0 = \lambda\left(v_z(z) - w(z,\xi) + \textstyle{\inf_{\hat \xi}w(z, \hat \xi)}\right) + o(\lambda).
$$
In the ansatz \eqref{eq:ansatz}, the function $w$ is multiplied by a higher-order $\lambda$ term. Therefore, its value at a particular point is irrelevant at the leading order $\lambda^{1/2}$ and we may assume $w(z,0) = 0$. As a result, we expect 
$\inf_{\hat \xi}w(z, \hat \xi) = w(z,0) = 0$, because a zero deviation $\xi=0$ from the frictionless position should lead to the smallest utility loss. Consequently, the leading-order dynamic programming equation outside the no-trade region reads as 
\begin{equation}\label{eq:dpetrade}
0=v_z(z)-w(z,\xi).
\end{equation}
Note that this derivation remains valid for several risky assets.

\subsubsection*{Corrector Equations}
Together with \eqref{eq:drift2}, \eqref{eq:dpetrade} shows that -- at the leading order $\lambda^{1/2}$ -- the Dynamic Programming Equation \eqref{eq:dpe} can be written as
\begin{equation}\label{eq:dpeapprox}
0=\max\left\{\scr{A}u(z) +\frac{\sigma^2}{2}\xi^2 v_{zz}(z) -\frac{\sigma^2}{2}\theta^2(z)(1-\theta_z(z))^2 w_{\xi\xi}(z,\xi),w(z,\xi)-v_z(z)\right\},
\end{equation}
where we set
\begin{equation}\label{eq:A}
\scr{A}u(z):=\beta u(z) -\scr{L}_0 u(z) + \kappa(z)u_z(z).
\end{equation}
To solve \eqref{eq:dpeapprox}, first treat the $z$-variable as constant and solve \eqref{eq:dpeapprox} as a function of $\xi$ only:
$$0=\max\left\{\frac{\sigma^2}{2} \xi^2 v_{zz}(z) -\frac{\sigma^2}{2}\theta^2(z)(1-\theta_z(z))^2 w_{\xi\xi}(z,\xi)+a(z),w(z,\xi)-v_z(z)\right\},$$
for some $a(z)$ that only depends on $z$ but not on $\xi$. Then, take $a(z)$ as given and solve for the function $u$ of $z$:
$$\scr{A}u(z)=a(z).$$
If both of these ``corrector equations'' are satisfied, \eqref{eq:dpeapprox} evidently holds as well. For several risky assets, the corresponding analogues read as follows:

\begin{mydef}[Corrector Equations]\label{def:correct}
For a given $z>0$, the \emph{first corrector equation} for the unknown pair $(a(z),
 w(z,\cdot)) \in \R_+ \times C^2(\R_+)$ is
 \begin{equation}
\label{e.cw}
 \max \Big\{ -\frac{|\sigma^{\top} \xi|^2}{2}\ (- v_{zz}(z))
 -\frac{1}{2}\mbox{\rm Tr}\left[\alpha(z)\alpha(z)^{\top}w_{\xi\xi}\right]
              +a(z),\,
 w(z,\xi) - v_{z}(z)
 \Big\}
 \;=\;
 0, \quad \forall \xi \in \mathbb{R}^d,
 \end{equation}
together with the normalization $w(z,0)=0$, where 
$$
\alpha(z):= (I_d-\theta_z(z) \1_d^T) \mathrm{diag}[\theta(z)]\sigma.
$$  
The \textit{second corrector equation} uses the function $a(z)$ from the first corrector equation
and is a simple linear equation for the function $u : \R_+
 \to \R$:
 \begin{equation}
\label{e.au}
 \Ac u(z) = a(z),
 \qquad
\forall \  z \in \R_+,
 \end{equation}
where $\Ac$ -- defined in \eqref{eq:A} and \eqref{eq:L0} -- is the infinitesimal generator of the
 optimal wealth process for the frictionless problem.
\end{mydef}

\begin{myrek}
\label{r.ergodic}
{\rm{
As for proportional costs \cite[Remark 3.3]{soner.touzi.12},
the first corrector equation is the dynamic programming equation 
of an ergodic control problem. Indeed, 
 for fixed $z$ and
 for an increasing sequence of stopping times 
 $\tau = (\tau_k)_{k\in \mathbb{N}}$ 
 and impulses $m = (m_k)_{k\in \mathbb{N}} \in \R^d$, 
 one defines the cost functional 
 by
$$
J(z,m, \tau) := v_z(z)\limsup_{T\to \infty}\frac{1}{T}\  
\E\left[\int_0^T \frac{(-v_{zz}(z))}{2 v_z(z)}|\sigma^{\top}
\xi_s|^2 ds + \sum_{k=1}^\infty 1_{\{\tau_k\leq T\}} \right],
$$
where the state process $\xi$ is given by
$$
 \xi^i_t
 =  \xi^i_0
 +\sum_{j=1}^d {\alpha}^{i,j}(z) B^j_t +
  \sum_{k=1}^\infty m_k\ 1_{\{\tau_k\leq t\}},
  \quad t \ge 0, \ i=1,\ldots,d,
  $$
  with a $d$-dimensional standard Brownian motion $B$. 

 The structure of the above problem implies that
 the optimal strategy is decreed through a region
 $\Cc$ enclosing the origin. The optimal stopping
 times are the hitting times of $\xi$ to the boundary of $\Cc$.
When $\xi$ hits $\partial \Cc$, it is optimal to move it to the origin. Hence, the optimal
stopping
 times $(\tau_k)$ are the hitting times of $\xi$ of the boundary of $\Cc$
 and $m_k= - \xi_{\tau_k^-}$, so that $\xi_{\tau_k}=0$ for each
 $k=1,2, \ldots$ Put differently, the region $\Cc$ provides
 the asymptotic shape of the no-trade region.
 In the power and log utility case, it is an ellipsoid as in Figure \ref{fig:corr}.  
 
The function $a$ is the optimal value,
$$
a(z):= \inf_{(\tau, m)}J(z,m,\tau).
$$
Then, the Feynman-Kac formula, for 
the linear equation $\scr{A}u=a$ for $u$ implies
$$
u(z)= \E \left[
\int_0^\infty\ e^{-\beta t}\ a(Z^{m,z}_t) dt\right],
$$
where $Z^{m,z}$ is the optimal wealth process for the 
frictionless Merton problem with initial value $Z^{m,z}_0=z$.
}}
\end{myrek}

\subsection{Solution in One Dimension}\label{sec:sol1}

If there is only a single risky asset ($d=1$), the asymptotically optimal no-trade region is an interval $\{z: |\xi| \leq \xi_0(z)\}$. The first corrector equation can then be readily solved explicitly by imposing smooth pasting at the boundaries, similarly as for proportional transaction costs \cite{soner.touzi.12}. Matching values and first derivatives across the trading boundaries $\pm \xi_0(z)$ leads to two conditions for a symmetric function $w(z,\cdot)$, in addition to the actual optimality equation in the interior of the no-trade region. Thus, the lowest order polynomials in $\xi$ capable of fulfilling these requirements are of order four. Since we have imposed $w(z,0)=0$, this motivates the ansatz
$$
w(z,\xi)=
\begin{cases} 
A(z)\xi^2-B(z)\xi^4, &|\xi| \leq \xi_0(z),\\ 
v_z(z), & |\xi| \geq \xi_0(z).
\end{cases}
$$ 
Inside the no-trade region, inserting this ansatz into the first corrector equation \eqref{e.cw} gives
$$0=a(z)+\frac{\sigma^2}{2}\xi^2 v_{zz}(z)- \alpha^2(z)A(z) + 6 \alpha^2(z) B(z) \xi^2,$$
where $\alpha(z):= \sigma \theta(z)(1-\theta_z(z))$ as in Definition \ref{def:correct} above. Since this equation should be satisfied for any value of $\xi$, comparison of coefficients yields
\begin{equation}\label{eq:coeff}
B(z)=\frac{\sigma^2}{12}\frac{-v_{zz}(z)}{\alpha^2(z)}, \quad A(z)=\frac{a(z)}{\alpha^2(z)}.
\end{equation}
Next, the smooth pasting condition $0= 2 A(z) \xi_0(z)-4 B(z) \xi_0^3(z)$ at the trading boundary $\xi=\xi_0(z)$ implies
\begin{equation}\label{eq:bound1}
\xi_0^2(z)=\frac{A(z)}{2B(z)}.
\end{equation}
Finally, the value matching conditions $v_z(z)=A(z)\xi_0^2(z)-B(z)\xi_0^4(z)$ at $\xi=\xi_0(z)$ give
$$v_z(z)=\frac{A^2(z)}{4B(z)}=-\frac{3a^2(z)}{\alpha^2(z)\sigma^2 v_{zz}(z)},$$
and in turn
\begin{equation}\label{eq:az}
a(z)=v_z(z) \alpha(z) \sigma \sqrt{-\frac{v_{zz}(z)}{3v_z(z)}}.
\end{equation}
In view of \eqref{eq:bound1} and \eqref{eq:coeff}, the optimal trading boundaries are therefore determined as 
\begin{equation}\label{eq:dev}
\xi_0(z)=\left(\frac{12}{-v_{zz}(z)/v_z(z)} \theta(z)^2 (1-\theta_z(z))^2\right)^{1/4}.
\end{equation}
For utilities with constant relative risk aversion $\gamma>0$, the optimal frictionless risky position is $\theta(z)=\pi_m z$, so that that the corresponding trading boundaries are given by
$$\xi_0(z)=\left(\frac{12}{\gamma} \pi_m^2 (1-\pi_m)^2 z^3 \right)^{1/4}.$$
For the maximal deviations of the risky weight from the frictionless target, this yields the formulas from Theorem \ref{thm:strategy}:
$$\pi_0(z)=\frac{\lambda^{1/4}\xi_0(z)}{z}= \left(\frac{12}{\gamma}\pi_m^2 (1-\pi_m)^2 \frac{\lambda}{z} \right)^{1/4}.$$
With constant relative risk aversion, the homotheticity of the value function \eqref{eq:valfun0} and \eqref{eq:az} imply that the second corrector equation $\scr{A}u(z)=a(z)$ simplifies to 
$$\beta u(z) -rzu_z-\frac{(\mu-r)^2}{\gamma\sigma^2} z u_z(z) -\frac{(\mu-r)^2}{2\gamma^2\sigma^2} z^2 u_{zz}(z)+c_m z u_z=\sqrt{\frac{\gamma}{3}}c_m^{-\gamma} \sigma^2 \pi_m(1-\pi_m) z^{1/2-\gamma},$$
which is solved by
$$u(z)=u_0 z^{1/2-\gamma} \quad \mbox{with } u_0=\sigma^2 \left(\frac{\gamma}{3} \pi_m^2(1-\pi_m)^2\right)^{1/2} \frac{c_m(\gamma)^{-\gamma}}{c_m(2\gamma)}.$$
This is the formula from Theorem \ref{t.main}.

\subsection{Solution in Higher Dimensions}\label{sec:multi}

Let us now turn to the solution of the corrector equations for multiple risky assets. To ease the already heavy notation, we restrict ourselves to utilities $U_\gamma$ with constant relative risk aversion $\gamma>0$ here. Then, we can rescale the corrector equation to obtain a version which is independent of the wealth variable $z$. Indeed, let 
$$
\rho = z^{-3/4}\xi
$$
so that, setting 
$$v_0=c_m^{-\gamma},$$
we obtain
$$
w(z,\xi) = v_z(z)W(z^{-3/4}\xi) = v_0z^{-\gamma}W(\rho),\quad a(z) = a_0 z^{1/2-\gamma} > 0,
$$
for some constant $a_0>0$ and a function $W(\rho)$ to be determined. We also introduce the matrices 
$$
A:= z^{-2}\alpha(z)\alpha(z)^{\top},\quad \Sigma:=\sigma\sigma^{\top}.
$$
Then, a direct computation shows
\begin{align*}
|\sigma^{\top}\xi|^2v_{zz}(z) =& (-\Sigma \rho\cdot \rho)\frac{v_0 z^{1/2-\gamma}}{\gamma},\\
\mbox{Tr}\left[\alpha(z)\alpha(z)^{\top}w_{\xi\xi}(z,\xi)\right]  = & \mbox{Tr}[A\ W_{\rho\rho}(\rho)](v_0z^{1/2-\gamma}).
\end{align*}
We use the notation $A:B:=\mbox{Tr}[AB]$ to rewrite the corrector equation. The resulting rescaled equation is for the pair $(W(\cdot), a_0)$, with independent variable $\rho\in \mathbb{R}^d:$
\begin{equation}\label{e.homothetic_corrector}
\max\left\{-\frac12 \Sigma\rho\cdot\rho - \frac12 [A : W_{\rho\rho}(\rho)] + a_0\ ;\ -1 + W(\rho)   \right\} = 0,
\end{equation}
together with the normalization $W(0) = 0.$ Following Atkinson and Wilmott \cite{atkinson.wilmott.95}, we postulate a solution of the form 
$$
W^*(\rho) = 1-(M\rho\cdot\rho - 1)^2,
$$
for a symmetric matrix $M$ to be computed. Then, 
$$
W_{\rho\rho}^*(\rho) = -4(M\rho\cdot\rho - 1)M - 8M\rho\otimes M\rho.
$$
Hence:
\begin{align*}
-\frac12 [A : W^*_{\rho\rho}(\rho)]  = & 2(M\rho\cdot\rho - 1)[A : M] + 4MAM\rho\cdot \rho\\
 = & (2M[A : M] + 4MAM)\rho\cdot\rho - A : M\\
 = & \frac12 \Sigma \rho\cdot\rho - a_0,
\end{align*}
provided that $a_0 = A : M$ and $M$ solves the algebraic Ricatti equation
\begin{equation}\label{e.ricatti}
4M[A : M] + 8MAM = \Sigma.
\end{equation}
Remarkably, this is exactly equation (3.7) obtained by Atkinson and Wilmott \cite{atkinson.wilmott.95} in their asymptotic analysis of the Morton and Pliska model \cite{morton.pliska.95} with trading costs equal to a constant fraction of the investors' current wealth. Atkinson and Wilmott \cite{atkinson.wilmott.95} argue that one may take $A$ to be the identity without any loss of generality by transforming to a coordinate system in which the second order operator is the Laplacian. For the convenience of the reader, we provide this transformation here: since $A$ is symmetric positive definite by Assumption \eqref{eq:posdef}, there is a unitary matrix $O\in \R^{d\times d}$ for which 
\[OAO^{\top} = \mathrm{diag}[\zeta_i],\]
where $\zeta_1,\zeta_2,\ldots, \zeta_d$ denote the eigenvalues of $A$. Setting
\begin{align*}
\tilde M  :=& \mathrm{diag}[\zeta_i^{1/2}] O M O^{\top} \mathrm{diag}[\zeta_i^{1/2}],\\
\tilde \Sigma :=& \mathrm{diag}[\zeta_i^{1/2}] O \Sigma O^{\top} \mathrm{diag}[\zeta_i^{1/2}],
\end{align*}
Equation \eqref{e.ricatti} becomes 
\begin{equation}\label{e.ricatti2}
4\tilde M\mbox{Tr}[\tilde M] + 8\tilde M^2 = \tilde \Sigma. 
\end{equation}
Using that $\tilde M$ and $\tilde \Sigma$ have the same eigenvectors, Atkinson and Wilmott (see (3.8-3.11) in \cite{atkinson.wilmott.95}) obtain simple algebraic equations for the eigenvalues of $\tilde M$, thus determining $M$ up to the above coordinate transformation. In summary, given $A,\Sigma$ positive definite, there exists a positive definite solution $M$ of \eqref{e.ricatti}. Then, the following function solves the corrector equation \eqref{e.cw}:
$$
W(\rho):= \left\{ \begin{array}{lc}
1- (\rho^\top M\rho - 1)^2, & \mbox{ for }\rho\in \scr{J},\\
1, & \mbox{ for }\rho\notin \scr{J},
\end{array} \right. 
$$
where $\scr{J}$ is the following ellipsoid around zero:
$$
\scr{J}:=\{\rho\in \mathbb{R}^d\ :\ \rho^\top M\rho  < 1   \}. 
$$
Reverting to the original variables, it follows that the asymptotically optimal no-trade region should be given by
$$\mbox{NT}^{\lambda} = \left\{ (x,y) \in \mathbb{R}^{d+1}: \frac{y}{x+y \cdot \1_d}   \in  \pi_m + \frac{\lambda^{1/4}}{(x+y \cdot \1_d)^{1/4}}  \scr{J}\right\},$$ 
in accordance with Theorem \ref{thm:strategy}.

\begin{myrek}\label{r.g}
\rm{
For each $z >0,$ define 
\[\scr{G}(z):= \left\{\xi\in \R^d\ :\  \xi\in z^{3/4} \scr{J}  \right\},\] 
so that, given, $(z,\xi)=\left(x+ y \cdot \1_d, \frac{y-\pi_m (x+ y\cdot \1_d)}{\eps}\right) \leftrightarrow(x,y)\in K_{\lambda},$ one has $(x,y)\in NT^{\lambda}$ if and only if $\xi\in \scr{G}(z)$.
}
\end{myrek}

\section{Proofs}

In the sequel, we turn the above heuristics into rigorous proofs of our main results, Theorems \ref{t.main} and \ref{thm:strategy}, using the general methodology developed by Barles and Perthame \cite{barles.perthame.87} and Evans \cite{evans.89} in the context of viscosity solutions. To ease notation by avoiding fractional powers, we write
$$\lambda = \eps^4,$$
and, with a slight abuse of notation, use sub- or superscript $\eps$ to refer to objects pertaining to the transaction cost problem. For instance, $v^{\eps}$ refers to $v^{\lambda}$, $K_{\eps}$ to $K_{\lambda},$ et cetera.

To establish the expansion of the value function asserted in Theorem \ref{t.main}, we need to show that
 \[u^{\eps}(x,y) = \frac{v(z) - v^{\eps}(x,y)}{\eps^2}\]
is locally uniformly bounded from above as $\eps\to 0$. To this end, define the \emph{relaxed semi-limits}
\be \label{e.relaxed}
u_*(x_0,y_0) = \liminf_{\stackrel{(\eps,x,y)\to (0,x_0,y_0)}{(x,y)\in K_{\eps}}} u^{\ep}(x,y),
& &u^*(x_0,y_0)= \limsup_{\stackrel{(\eps,x,y)\to (0,x_0,y_0)}{(x,y)\in K_{\eps}}} u^{\ep}(x,y).
\ee
Their existence is guaranteed by the straightforward lower bound $u^{\eps}\geq 0$, as well as the locally uniform upper bound provided in Theorem \ref{t.upperbound}.  Establishing the latter involves the explicit construction of a particular trading strategy and is addressed first. We then show in Sections \ref{s.subsolution} and \ref{s.supersolution} that the relaxed semi-limits $u^*, u_*$ are viscosity sub- and super-solutions, respectively, of the Second Corrector Equation \reff{e.au}. Combined with the Comparison Theorem \ref{t.comparison} for the second corrector equation provided in Section \ref{s.comparison}, this in turn yields that $u^*\le u_*$. Since the opposite inequality is satisfied by definition, it follows that $u=u^*=u_*$ is the unique solution of the Second Corrector Equation \reff{e.au}. As a consequence, $u^{\eps}\to u$ locally uniformly, verifying the asymptotic expansion of the value function. With the latter at hand, we can in turn verify that the policy from Theorem \ref{thm:strategy} is indeed almost optimal for small costs (cf.\ Section \ref{s.mainproof}).

\subsection{Existence of the Relaxed Semi-Limits}

\subsubsection*{Locally uniform upper bound of $u^{\eps}$}

In this section we show that $u^{\eps}(x,y) = \eps^{-2}(v(z) - v^{\eps}(x,y))$ is locally uniformly bounded from above as $\eps\to 0$:

\begin{mythm}\label{t.upperbound}
Given any $x_0,y_0$ with $x_0+y_0\cdot \1_d>0$, there exists $\eps_0>0$ and $r_0=r_0(x_0,y_0)>0$ such that 
\begin{equation}\label{e.upperbound}\sup\{u^{\eps}(x,y):\ (x,y)\in B_{r_0}(x_0,y_0), \eps\in(0,\eps_0] \}<\infty.
\end{equation}
\end{mythm}
Theorem \ref{t.upperbound} is an immediate corollary of Theorem \ref{t.bounds} below. To prove the latter, we construct an investment-consumption policy which gives rise to a suitable upper bound. The construction necessitates some technical estimates. The reader can simply read the definition of the strategy and proceed directly to the proof of Theorem \ref{t.bounds} in order to view the thread of the argument.

\paragraph*{Strategy up to a Stopping Time $\theta$}
Given an initial portfolio allocation $(X_{0^-},Y_{0^-})\in K_{\eps}$, use the trading strategy from Theorem \ref{thm:strategy}, corresponding to the no-trade region $\mbox{NT}^{\eps}$, from time $0$ until a stopping time $\theta$ to be defined below. More specifically, let $(\tau_1,\tau_2,\ldots; m_1,m_2,\ldots)$, where $\tau_i$ is the $i$-th time the portfolio process hits the boundary $\partial\mbox{NT}^{\eps}$ of the no-trade region. The corresponding reallocations $m_1,m_2,\ldots$ are chosen so that, after taking into account transaction costs, the portfolio process is at the frictionless Merton proportions:
\[\frac{Y^{\eps}_{\tau_i}}{X^{\eps}_{\tau_i}+Y^{\eps}_{\tau_i}\cdot \1_d}= \pi_m. \]
Until $\theta$, the investor consumes the optimal frictionless proportion of her current wealth,
\[c_{t} = c_m Z_{t}^{\eps},\quad \forall t\leq \theta,\]
so that her wealth process is governed the following stochastic differential equations until time $\theta$:
\begin{eqnarray}\label{e.trading}
X^{\eps}_t & =& X_{0^-}+ \int_0^t(rX^{\eps}_s - c_s)ds - \sum_{k=1}^{\infty}\left(\eps^4 + \sum_{j=1}^d m^j_k\right) 1_{\{\tau_k \leq t\}},\nonumber\\
Y^{\eps}_t & = & Y_{0^-}+ \int_0^tY^{\eps}_s\frac{dS_s}{S_s} +\sum_{k=1}^{\infty}m_k 1_{\{\tau_k \leq t\}}.
\end{eqnarray}
The stopping time $\theta$ must be chosen so that the investor's position remains solvent at all times, $(X_t^{\eps},Y_t^{\eps})\in K_{\eps},$ $\forall  t\leq \theta $, $\P$-almost surely. Therefore, we use the first time the investor's wealth falls below some threshold, which needs to be large enough to permit the execution of a final liquidating bulk trade. 

\paragraph{At Time $\theta$ and Beyond}

Define $\theta = \theta^{\eta,\eps}$ to be the exit time of the portfolio process from the set
\begin{eqnarray*}
K^{\eta,\eps}:= \{(z,\xi)\in \R_+\times \mathbb{R}^d&:&  \mbox{either}\quad z > (\eta+1)\eps^4\mbox{ and } \xi \in \R^d\\
& & \mbox{or } z\in (\eta\eps^4, (\eta+1)\ep^4]\mbox{ and }\xi \in \scr{G}(z) \}.
\end{eqnarray*}
Within $K^{\eta, \eps},$ the above policy is used and the portfolio process follows \eqref{e.trading}. At time $\theta,$ the investor liquidates all risky assets, leading to a safe position of least $(\eta-1)\eps^4$. Afterwards, she consumes at half the interest rate, thereby remaining solvent forever. The resulting portfolio process satisfies the following deterministic integral equation with stochastic initial data:
\begin{equation}\label{e.deterministic}
X^{\eps}_{\theta+t} =  X^{\ep}_{\theta} + \int_0^t\frac{r}{2} X^{\eps}_{\theta+s}dt, \quad  Y^{\eps}_{\theta+t}  =  0, \quad t \geq 0.
\end{equation}
Let $(X, Y)^{\eta,\ep}_t,$ $t\geq 0,$ be the portfolio process produced by concatenating the controlled stochastic process \eqref{e.trading} and the deterministic process \eqref{e.deterministic} at time $\theta$.

\begin{myrek}\label{r.deterministic}
\rm{
For any $\eta > 1$, the optimal value $v^{\eps}(\eta\eps^4, \xi)$ must be greater than or equal to the utility obtained from the immediate liquidation of all risky assets and then running the deterministic policy \eqref{e.deterministic}. Since the latter can be computed explicitly, this provides a crude lower bound for $v^{\eps}(\eta\eps^4, \xi)$. 

To see this, suppose the investor's wealth after the liquidating trade at time $\theta$ is given by $X_{\theta}^{\eps} \geq (\eta - 1)\eps^4$. Then,  $X_{\theta+t}^{\ep} \geq (\eta-1)\eps^4e^{\frac{r}{2}t}$. For power utilities ($0<\gamma \neq 1$), this yields the lower bound
\begin{equation*}
v^{\eps}(\eta\eps^4, \xi)  \geq  \int_{0}^{\infty}e^{-\beta t}\frac{(\eta-1)^{1-\gamma}\eps^{4-4\gamma}e^{\frac{r}{2}(1-\gamma)t}}{1-\gamma}dt=  \frac{(\eta-1)^{1-\gamma}}{(1-\gamma)(\beta -\frac{r}{2}(1-\gamma))}\eps^{4-4\gamma}.
\end{equation*}
The corresponding result for logarithmic utility ($\gamma=1$) is
\begin{equation}\label{eq:boundlog}
v^{\eps}(\eta\eps^4, \xi)  \geq  \int_{0}^{\infty}e^{-\beta t}\log[(\eta-1)\eps^4e^{rt/2}]  dt=   \frac{\log[(\eta-1)\eps^4]}{\beta} + \frac{r}{2\beta^2}.
\end{equation}
}
\end{myrek}

\paragraph*{Constructing a Candidate Lower Bound}

For given $\eps,\delta, C > 0$, define the function 
\[
V^{\eps,\delta}_C(z,\xi) = v(z) - \eps^2Cu(z) - \eps^4(1+\delta)w(z,\xi).
\]
We now establish a series of technical lemmata. These will be used in the proof of Theorem \ref{t.bounds} to verify that -- asymptotically -- $V^{\eps,\delta}_C$ is dominated by the value function $v^\eps$ in the no-trade region for an appropriate choice of the parameters $C$ and $\delta$.

\begin{mylemma}\label{l.eta-level-bound}
Let $\eta > 1$ be given. There exists $C_{\eta}>0,$ independent of $\eps,$ such that for all $\bar{z} \in [\eta\eps^4, (\eta+1)\eps^4]$ we have
\[v^{\eps}(\bar{z},\xi) \geq V^{\eps,\delta}_{C_{\eta}}(\bar{z},\xi),\quad\mbox{for all }\xi\in \mathbb{R}. \]
\end{mylemma}

\proof
We only consider power utilities ($\gamma \neq 1$); the case of logarithmic utility can be treated similarly. First, notice that since the term $-\eps^4(1+\delta)w(\bar{z},\xi)$ is always negative it can be ignored. Write $\bar z = (\eta+\bar \lambda)\eps^4$, for some $\bar \lambda \in [0,1].$ Using the estimates from Remark \ref{r.deterministic}, the goal is to find a sufficiently large $C_{\eta}$ so that
\begin{align*}
\frac{(\eta - 1 +\bar \lambda )^{1-\gamma}}{(1-\gamma)(\beta - \frac{r}{2}(1-\gamma))}\eps^{4-4\gamma} &\geq v(\bar z) - C_{\eta}\eps^2 u(\bar z)\\
&= (\eta+\bar\lambda)^{1-\gamma}\left(\frac{v_0}{1-\gamma} - C_{\eta} (\eta+\bar \lambda)^{-1/2} u_0\right)\eps^{4-4\gamma},
\end{align*}
for all $\bar \lambda \in [0,1].$ This follows by observing that one can take
\begin{align}\label{e.C_eta}
C_{\eta} := \mbox{constant}\times \sqrt{\eta},
\end{align} 
for a large enough positive constant which only depends on the model and preference parameters ($\mu,r,\sigma,\gamma,\beta$) but is independent of $\eps$ and $\eta$. 
\qed

\begin{mylemma}\label{l.jumps}
There exists $\delta > 0$ such that, for all $\eta$ sufficiently large, there is $\eps_0 = \eps_0(\eta,\gamma)>0$ such that
\[V^{\eps,\delta}_{C_{\eta}}(z-\eps^4,0) - V^{\eps,\delta}_{C_{\eta}}(z,\bar\xi)\geq 0,\quad \forall \eps\in (0,\eps_0],\  z\geq \eta\eps^4,\  \bar\xi\in \partial\scr{G}(z) . \]
\end{mylemma}

\proof 
Recall that, by definition, the corrector $w$ satisfies $w(\cdot,0) = 0$ as well as $w(z,\bar\xi) = v_z(z)$ for $\bar\xi\in \partial\scr{G}(z).$

First consider the case of power utility. Taylor expansion and evaluation at $z=\eta \eps^4$ yields
\begin{align}\label{e.jumps}
V^{\eps,\delta}_{C_{\eta}}&(z-\eps^4,0) - V^{\eps,\delta}_{C_{\eta}}(z,\bar\xi)\nonumber\\
&=  v(z-\eps^4) - v(z) - \eps^2C_{\eta}(u(z-\eps ^4)-u(z))+ \eps^4(1+\delta)v_z(z) \nonumber\\
&= \delta \eps^4 v_z(z)  + \eps^6 C_{\eta} u'(z) - \eps^8 v_{zz}(\tilde z_1) + \eps^{10} C_{\eta}u''(\tilde z_2)\nonumber\\
& = \eps^{4-4\gamma}\left(v_0\left(\delta \eta^{-\gamma} + \gamma\tilde \eta_1^{-1-\gamma}\right) +C_{\eta}u_0(1/2-\gamma)\left(\eta^{-1/2-\gamma} + (1/2+\gamma)\tilde \eta_2^{-3/2-\gamma}\right)    \right),
\end{align}
where the points $\tilde z_1, \tilde z_2\in [z-\eps^4,z]$ are determined by the Taylor remainders of $v$ and $u$, respectively, and $\tilde \eta_1, \tilde \eta_2\in [\eta-1, \eta]$ satisfy $\tilde z_i = \tilde \eta_i \eps^{4}.$ Considering expression \eqref{e.jumps} as a function of $\eta$, the dominant term is of the order $O(\eta^{-\gamma}).$ Since $C_{\eta} = C \sqrt{\eta}$ where $C$ only depends on the model parameters ($\mu,r,\sigma,\gamma,\beta$), the term $C_{\eta}\eta^{-1/2-\gamma}$ also contributes at the order $O(\eta^{-\gamma})$. Consequently, choosing
 \begin{align}\label{e.delta}
 \delta > \frac{Cu_0|1/2-\gamma|}{v_0},
\end{align} 
ensures that the leading-order coefficient is positive, independent of $\eta$. For sufficiently large $\eta$, the assertion follows.

In the case of logarithmic utility ($\gamma=1$), the argument is the same because the expressions involved are all power type functions. The same choice \eqref{e.delta} of $\delta$ works as well.
\qed

\begin{mylemma}\label{l.utility-domination}
For sufficiently large $\eta$ there exists $\eps_0=\eps_0(\eta,\gamma)>0$ such that
\[\beta V^{\eps,\delta}_{C_{\eta}}(z,\xi) - \scr{L} V^{\eps,\delta}_{C_{\eta}}(z,\xi) \leq U(c_m z),\quad \forall \eps \in (0,\eps_0],\ z\geq \eta\eps^4,\ \xi\in \scr{G}(z), \]
where $\delta$ is given by \eqref{e.delta}.
\end{mylemma}

\proof We consider only the power utility case as the argument also works mutatis mutandis for logarithmic utility. To ease notation, we write $V^{\eps}$ instead of $V^{\eps,\delta}_{C_{\eta}}$. Throughout the proof, $(x,y)\in K_{\eps}$ satisfies $z = x+y\cdot\1_d = \eta\eps^4.$ Decompose 
\begin{align*}
\beta V^{\eps}(x,y) - &\scr{L} V^{\eps}(x,y) \\ &=\underbrace{(\beta v(z) - \scr{L} v(z))}_{=:I_1(z)} - \underbrace{\eps^2 C_{\eta} (\beta u(z) - \scr{L} u(z))}_{=:I_2(z)} - \underbrace{\eps^4(1+\delta)(\beta w(z,\xi) - \scr{L} w(z,\xi))}_{=:I_3(z)}.
\end{align*}
We analyze the asymptotic properties in $\eta$ of each of the terms $I_1, I_2,$ and $I_3$:
\begin{align*}
I_1(z) & =  -\frac12 \eps^2|\sigma^{\top}\xi|^2 v_{zz}(z) + \tilde{U}(v_z(z))\\
& =  -\frac12 \eps^2|\sigma^{\top}\xi|^2 v_{zz}(z) + {U}(c_m z) - c_mzv_z(z)\\
& \leq  -\frac12 \eps^2|\sigma^{\top}\xi|^2 v_{zz}(z) + {U}(c_m z) - \eps^2 C_{\eta} c_mzu_z(z)\\
&\leq \eps^{4-4\gamma}O(\eta^{1/2-\gamma}) + {U}(c_m z) - \eps^2 C_{\eta} c_mzu_z(z).
\end{align*}
Here, we used that $v_z(z)\geq \eps^2 C_{\eta}u_z(z)$ if $\eta$ is sufficiently large. The estimates in Remark \ref{r.estimates}, and the fact that $C_{\eta}$ is of order $O(\sqrt{\eta})$ (cf. Equation \eqref{e.C_eta}) give
\begin{align*}\label{e.I_2}
I_2(z) + \eps^2C_{\eta}c_mzu_z(z) 
&= C_{\eta}\left(\eps^2\Ac u(z)  - \eps^3\xi\cdot(\mu-r\1_d) u_z(z)
- \eps^3\left(\frac{1}{2}\eps|\sigma^{\top}\xi|^2 - \sigma^{\top} \xi\cdot\sigma^{\top} \pi_mz\right)u_{zz}(z)\right)\nonumber\\
 &\geq C_{\eta}\left(\eps^2a_0z^{1/2-\gamma} - K(\eps^3z^{1/4-\gamma} + \eps^4z^{-\gamma})\right)\\
 & = \eps^{4-4\gamma}O(\eta^{1-\gamma}),
\end{align*}
where $a_0=a(z)/z^{1/2-\gamma}$. Hence, this term is positive for sufficiently large $\eta$.  Finally, by Remark \ref{r.estimates}, we have
\begin{align*}
|I_3(z)| & \leq  (1+\delta)K(\eps^2z^{1/2-\gamma} + \eps^3z^{1/4-\gamma}+ \eps^4z^{-\gamma}+ \eps^5z^{-1/4-\gamma}  + \eps^6z^{-1/2-\gamma})\\
& = \eps^{4-4\gamma}O(\eta^{1/2-\gamma}),
\end{align*}
again for all sufficiently large $\eta$. In summary:
\begin{align*}
\beta V^{\eps}(x,y) - \scr{L} V^{\eps}(x,y)&= I_1(z) -I_2(z) - I_3(z)\\
&\leq U(c_m z) + \eps^{4-4\gamma}(O(\eta^{1/2-\gamma}) - O(\eta^{1-\gamma}))\\
& \leq U(c_m z),
\end{align*}
for sufficiently large $\eta$. Equivalently, there exists some $\eta>1$ such that, for all $z\geq \eta\eps^4$ and $\xi\in \scr{G}(z)$: 
\begin{align*}
\beta V^{\eps}(z, \xi) - \scr{L}V^{\eps}(z,\xi) & \leq  U(c_m z),\quad \forall \eps\in (0,\eps_0].
\end{align*}
This completes the proof.
\qed\\


We now have all the ingredients to prove the main result of this section, which in turn yields Theorem \ref{t.upperbound}.

\begin{mythm}\label{t.bounds}
There are constants $C,\delta, \eps_0 > 0$ such that, for all $\eps\in (0,\eps_0]$:
\be\label{eq:t.bounds}
v(z)-\eps^2Cu(z) -\eps^4(1+\delta)w(z,\xi) & \leq & {v}^{\eps}(z,\xi),\quad \forall (z,\xi)\in K^{\eps}.
\ee
In particular,
\begin{equation}\label{eq:upperbound}
u^{\eps}(z,\xi) \leq Cu(z) + o(\eps),\quad \forall (z,\xi)\in K_{\eps},
\end{equation}
so that \eqref{e.upperbound} is satisfied.
\end{mythm}

\proof
Let $(x,y)\in K^{\eps}$ be given and let $\eta > 1$ be large enough so that all the previous lemmata are applicable. Without loss of generality, we may assume $x+y>\eta\eps^4$, since we are proving an asymptotic result. 

\textit{Step 1:} Let $(X_t,Y_t):=(X^{x,\eta,\ep}_t,Y^{y,\eta,\ep}_t)$ be the controlled portfolio process with dynamics \eqref{e.trading}, \eqref{e.deterministic}, which starts from the initial allocation $(x,y)$ and switches to deterministic consumption at half the interest rate at the first time $\theta:=\theta^{\eta}$ the total wealth $Z_t := X_t+Y_t$ falls to level $z=\eta\eps^4$. As before, write 
\[V^{\eps}(z,\xi):= V^{\eps,\delta_{\eta}}_{C_\eta}(z,\xi).\]
Recall that $C_{\eta}$ and $\delta_{\eta}$ are given by \eqref{e.C_eta} and \eqref{e.delta}, respectively.  
It\^o's formula yields
\begin{align*}
e^{-\beta\theta}V^{\eps}(X_{\theta},Y_{\theta}) =& V^{\eps}(x,y) - \int_0^{\theta}e^{-\beta s}(\beta V^{\eps}(X_s,Y_s) - \scr{L}V^{\eps}(X_s,Y_s))ds\\
&+\int_0^{\theta}e^{-\beta t}\mathbf{D}_y V^\eps(X_s,Y_s)^\top \sigma dW_s + \sum_{t \leq \theta}(V^{\eps}(Z_{t},\xi_{t}) - V^{\eps}(Z_{t-},\xi_{t-})).\nonumber
\end{align*}
Observe that the summation is at most countable and that, in view of Lemma \ref{l.jumps}, each summand satisfies
\[V^{\eps}(Z_{t},\xi_{t}) - V^{\eps}(Z_{t-},\xi_{t-})\geq 0.\]
Together with Lemma \ref{l.utility-domination}, this yields
\begin{align}\label{e.ito}
e^{-\beta\theta}V^{\eps}(X_{\theta},Y_{\theta}) \geq  V^{\eps}(x,y) - \int_0^{\theta}e^{-\beta s}U(c_mZ_s)ds
+\int_0^{\theta}e^{-\beta t}\mathbf{D}_y V^\eps(X_s,Y_s)^\top \sigma dW_s.
\end{align}

\textit{Step 2:} For any $(x',y')\in K^{\eps}$ with $0 < x'+y'\cdot \1_d \leq \eta\eps^4,$ let $\nu^{(x',y')}\in \Theta_{\eps}(x',y')$ be the strategy of \eqref{e.deterministic}, i.e., liquidation of all risky assets and then deterministic consumption at half the risk-free rate ad infinitum. According to Remark \ref{r.deterministic} and the proof of Lemma \ref{l.eta-level-bound}, \[v^{\eps}(X_{\theta},Y_{\theta}) \geq J(\nu^{(X_{\theta},Y_{\theta})}) \geq V^{\eps}(X_{\theta},Y_{\theta}),\quad \mbox{ on } \{\theta < \infty\},\] where \[J(\nu) :=\E \left[ \int_0^{\infty}e^{-\beta t}U(c_t)dt \right],\mbox { for any }\nu = (c,\tau,m).\] Let $\{\tau_n\}_{n\geq 0}$ be a localizing sequence of stopping times for the local martingale term in \eqref{e.ito} and set $\theta_n:=\theta\wedge\tau_n$. Assume for the moment that the family $\{e^{-\beta\theta_n}V^{\eps}((X,Y)_{\theta_n})\}_{n\geq 0}$ is uniformly integrable, and therefore it converges in expectation to its pointwise limit. Then, the same applies to the integral of the $dt$ term in \eqref{e.ito} by the dominated convergence theorem. Taking expectations in \eqref{e.ito}, sending $n\to \infty,$ and using these observations together with Lemmas \ref{l.eta-level-bound} and \ref{l.utility-domination} shows
\b*
V^{\eps}(x,y) &\leq& \mathbb{E}\left[\int_0^{\theta}e^{-\beta s}U(c_mZ_{s})ds + e^{-\beta\theta}V^{\eps}((X,Y)_{\theta}) \right] \\
& \leq & \mathbb{E}\left[\int_0^{\theta}e^{-\beta s}U(c_mZ_{s})ds +e^{-\beta\theta} \int_0^{\infty}e^{-\beta t}U(Z_{\theta}e^{rt/2})dt   \right]\\
&\leq & v^{\eps}(x,y),\quad \forall \eps\in (0,\eps_0].
\e*
As $x,y$ were arbitrary, the assertion \eqref{eq:t.bounds} follows. 

\textit{Step 3:} All that remains to show is that $\{e^{-\beta\theta_n}V^{\eps}((X,Y)_{\theta_n})\}_{n\geq 0}$ is uniformly integrable. Since the functions and domains are explicit, one can check that there is a constant $M > 0$, independent of $\eps,n$ such that \[|e^{-\beta\theta_n}V^{\eps}(X_{\theta_n},Y_{\theta_n})|\leq M|e^{-\beta\theta_n}v(Z_{\theta_n})|.\] Hence, it is sufficient to show that $\{e^{-\beta\theta_n}v(Z_{\theta_n})\}_{n\geq 0}$ is uniformly integrable. This will follow, for instance, if it is uniformly bounded in $L^{1+q}(\Omega)$ for some $q > 0.$ The interesting case is $0 < \gamma \leq 1$; otherwise $v(z)$ is bounded on the domain under consideration because the wealth process is bounded away from zero and the Merton value function is negative. We just show the power utility case; a similar argument applies for logarithmic utility. 

 Let $\tilde{Z}_t:=\tilde{Z}^{x,y,\eta, \eps}_t$ denote the same controlled wealth process, however, obtained by not deducting transaction costs or consumption. Evidently, $Z_{\tau}^{1-\gamma} \leq \tilde{Z}_{\tau}^{1-\gamma}$ almost surely, for any stopping time $\tau$. Moreover, for any $a,b>0$, we have
\begin{eqnarray*}
d[ e^{-a\beta \theta_n}(\tilde {Z}_{\theta_n})^b] & = e^{-a\beta \theta_n}\tilde{Z}_{\theta_n}^b\left\{(-a\beta + b(r + \pi_{t}\cdot(\mu - r \1_d) + \frac{b-1}{2}|\sigma^{\top}\pi_t|^2))dt + b\pi_t\cdot \sigma dW_{t}\right\},
\end{eqnarray*}
where $\pi_t:= Y_t/Z_t.$ When $a=1, b = 1-\gamma$, the drift term is maximized at the Merton proportion, $\pi_t\equiv \pi_m$, and moreover, by the finiteness criterion for the frictionless value function:
 \[-\beta + (1-\gamma)\left(r + \pi_m\cdot(\mu - r\1_d) - \frac{1}{2}|\sigma^{\top}\pi_m|^2\right)<0 .\] 
Taking $b=(1-\gamma)(1+q)$ and $a = (1+q)$ for sufficiently small $q>0$, the drift term is maximized at a vector $\pi^{a,b}$ arbitrarily close to $\pi_m$, for which
 \begin{eqnarray}\label{e.drift}A:=-a\beta + b\left(r + \pi^{a,b}\cdot(\mu - r\1_d) - \frac{1}{2}|\sigma^{\top}\pi^{a,b}|^2\right)<0. \end{eqnarray} 
 As a consequence: 
\begin{eqnarray*}
d[ e^{-a\beta \theta_n}(\tilde {Z}_{\theta_n})^b] & \leq e^{-a\beta \theta_n}\tilde{Z}_{\theta_n}^b\left\{(-a\beta + b(r + \pi^{a,b}\cdot(\mu - r \1_d) + \frac{b-1}{2}|\sigma^{\top}\pi^{a,b}|^2))dt +  b\pi_t\cdot \sigma dW_{t}\right\}.
\end{eqnarray*}
Taking expectations, passing to the limit over a localizing sequence of stopping times for the local martingale term, and applying Fatou's lemma, we obtain 
\begin{eqnarray*}
\|e^{-\beta \theta_n}(\tilde {Z}_{\theta_n})^{1-\gamma}\|_{L^{1+q}(\Omega)}^{1+q} = \E[ e^{-a\beta \theta_n}(\tilde {Z}_{\theta_n})^b] & \leq \tilde{Z}_0^b \E [\exp(A\theta_n)] \leq z^b,\quad \forall n\in \N.
\end{eqnarray*}
Hence, the family is uniformly bounded in $L^{1+q}(\Omega)$ and thus uniformly integrable as claimed. 
\qed\\

We conclude this section by establishing that the relaxed semi-limits $u_*,u^*$ only depend on total wealth and can be realized by restricting to limits on the Merton line.

\begin{mylemma}\label{l.relaxed}
For any $x_0+y_0\cdot\1_d > 0$, we have
\b* 
u_*(x_0,y_0) = \liminf_{\stackrel{(\eps,x,y)\to (0,x_0,y_0)}{(x,y)\in K_{\eps}}} u^{\ep}(z-\pi_m\cdot \1_d z ,\pi_m z),
& &u^*(x_0,y_0)= \limsup_{\stackrel{(\eps,x,y)\to (0,x_0,y_0)}{(x,y)\in K_{\eps}}} u^{\ep}(z-\pi_m\cdot \1_d z ,\pi_m z).
\e*
\end{mylemma}

\proof
Given $(x,y)\in K_{\eps}$, where $z = x+y\cdot\1_d > \eps^4$ without loss of generality, we observe that
\begin{eqnarray}\label{e.bounds_veps}
\inf_{{x'+y'\cdot\1_d = z+\eps^4}} v^{\eps}(x',y') \quad \geq\quad v^{\eps}(x,y) \quad \geq\quad \sup_{{x'+y'\cdot\1_d = z-\eps^4}} v^{\eps}(x',y').
\end{eqnarray}
Therefore,
\begin{eqnarray*}
\frac{v(z) - v^{\eps}(x,y)}{\eps^2} & \leq & \eps^{-2}\left(v(z -\eps^4)\quad - \sup_{x'+y'\cdot\1_d = z-\eps^4}v^{\eps}(x', y')\right) + \eps^2v_z(z-\eps^4)\\
& = & \inf_{x'+y'\cdot\1_d = z-\eps^4} u^{\eps}(x',y') + \eps^2v_z(z-\eps^4)
\end{eqnarray*}
and
\begin{eqnarray}\label{e.monotonicity}
\inf_{x'+y'\cdot\1_d = z-\eps^4} u^{\eps}(x',y') +\eps^2v_z(z-\eps^4) \geq u^{\eps}(x,y) \geq\hspace{0.1in} \sup_{x'+y'\cdot\1_d = z+\eps^4} u^{\eps}(x',y') - \eps^2v_z(z+\eps^4).
\end{eqnarray}
Let $(\eps_n,x_n,y_n)\to (0,x_0,y_0)$ be chosen so that we have $u^{\eps_n}(x_n,y_n)\to u_*(x_0,y_0).$ Setting $z_n' = x_n+y_n\cdot\1_d +\eps^4$ and using the previous observations, it follows that
\begin{eqnarray*}
u^{\eps_n}(x_n,y_n) &\geq& u^{\eps_n}(z_n'-\pi_m\cdot \1_d z_n' ,\pi_m z_n') - O(\eps^2).
\end{eqnarray*}
Taking $\liminf$ as $n\to \infty$ on both sides yields \[u_*(x_0,y_0) = u_*(z_0-\pi_m\cdot \1_d z_0 ,\pi_m z_0),\] where $z_0 = x_0 + y_0\cdot\1_d.$ The proof for $u^*$ is similar.
\qed\\
\begin{myrek}\label{r.envelopes}\rm{
For later use, observe that
\b*
u_*(x_0,y_0) = \liminf_{\stackrel{(\eps,x,y)\to (0,x_0,y_0)}{(x,y)\in K_{\eps}}} \underline {u}^{\ep}(x,y),& &
u^*(x_0,y_0) = \limsup_{\stackrel{(\eps,x,y)\to (0,x_0,y_0)}{(x,y)\in K_{\eps}}} \overline{u}^{\ep}(x,y),
\e*
where $\underline{u}^{\eps}, \overline{u}^{\eps}$ are the lower and upper semi-continuous envelopes of $u^{\eps}$, respectively. Moreover, \eqref{e.monotonicity} extends to the envelopes as follows:
\begin{align}\label{e.monotonicity2}
\overline{u}^{\eps}(x,y) &\leq \inf_{x'+y'\cdot\1_d = z-\eps^4} \overline{u}^{\eps}(x',y') + \eps^2v_z(z-\eps^4),\\
\underline{u}^{\eps}(x,y) &\geq \sup_{x'+y'\cdot\1_d = z+\eps^4} \underline{u}^{\eps}(x',y') - \eps^2v_z(z+\eps^4).\nonumber
\end{align}
}
\end{myrek}

\subsection{Viscosity Sub-Solution Property}
\label{s.subsolution}

\begin{mythm}\label{t.subsolution}
The function $u^*$ is a viscosity sub-solution of the Second Corrector Equation \eqref{e.au}.
\end{mythm}

\proof
Let $(z_0, \varphi)\in (0,\infty)\times C^2(\mathbb{R}_+)$ so that \[0=(u^*-\varphi)(z_0) > (u^*-\varphi)(z),\quad \forall z>0, z\neq z_0.\] To prove the assertion, we have to show 
\[\mathcal{A}\varphi(z_0)\leq a(z_0).\]

\textit{Step 1}: By Theorem \ref{t.bounds}, there exist $\eps_0$, $r > 0$ depending on $(x_0,y_0)$ so that \begin{align}\label{e.b*}b^* := \sup_{(x,y)\in B_{r}, \eps\in (0,\eps_0]}\overline{u}^{\eps}(x,y) < \infty,\quad B_r:=B_{r}(x_0,y_0).\end{align} The radius $r$ can taken small enough that $B_r$ does not intersect the line $z=0.$
By Lemma \ref{l.relaxed}, $u^*(z_0)$ can be achieved along a sequence $(z^{\eps}, 0)$ on the Merton line, i.e.,
 \[z^{\eps}\to z_0 \mbox{ and }\overline{u}^{\eps}(z^\eps,0)\to u^*(z_0), \mbox{ as }\eps\to 0.\] 
 Observe that \[\ell_{\eps}^*:= \overline{u}^{\eps}(z^{\eps})-\varphi(z^{\eps})\to 0\] and \[(x^{\eps},y^{\eps}):=(z^{\eps}-\pi_m\cdot\1_d z^{\eps}, \pi_m z^{\eps})\to (x_0,y_0):= (z_0 - \pi_m\cdot\1_d z_0, \pi_m z_0).\] Due to the strict maximality of $u^*-\varphi$ at $z_0$, each $z^{\eps}$ can be taken to be a maximizer of $\overline{u}^{\eps}(\cdot,0)-\varphi(\cdot)$ on $[z_0-r, z_0+r].$ For $\ep\in (0,\eps_0]$ and $\delta > 0$ set 
 \[{\psi}^{\ep,\delta}(z,\xi):= v(z)-\ep^2(\varphi(z) + \ell^{*}_{\ep}+ C(z-z^{\eps})^4)- \ep^4(1+\delta)w(z,\xi),\] 
 with $C>0$ to be chosen later. 

\textit{Step 2}: Now, we use the function $\psi^{\ep,\delta}$ to touch $\underline{v}^{\eps}$ from below near $(x_0,y_0)$. Set
\[
I^{\eps,\delta}(z,\xi)  :=  (\underline{v}^{\eps}-\psi^{\eps,\delta})(z,\xi).\]
Consider any point $(z,\xi)\leftrightarrow(x,y)\in B_r$. We have
\begin{align}\label{e.bigC}
\eps^{-2}I^{\eps}(z,\xi) & =  -\overline{u}^{\eps}(z,\xi) + \varphi(z)+\ell^*_{\ep} + C(z-z^{\ep})^4 +\eps^2(1+\delta)w(z,\xi)\nonumber\\
& \geq  -b^* + \varphi(z)+\ell^*_{\ep} + C(z-z^{\ep})^4.
\end{align}
Thus, $C>0$ can be chosen large enough to ensure that \eqref{e.bigC} is positive for all sufficiently small $\eps>0$ when $r>|z-z_0|>r/2$. 

Next, we show that $I^{\eps,\delta}(z,\xi)>0$ when $|z-z_0| < r$ and $\xi\notin \scr{G}(z)$ (recall the definition of $\scr{G}(z)$ in Remark \ref{r.g}). To this end, observe that by Taylor expansion, \eqref{e.monotonicity2}, and the maximizer property of $z^{\eps}$, we have 
\begin{align*}
\eps^{-2}I^{\eps,\delta}(z,\xi) & =  -\overline{u}^{\eps}(z,\xi) + \varphi(z) + \ell^*_{\ep} + C(z-z^{\ep})^4 + \eps^2(1+\delta)v_z(z)\\
& \geq  -\overline{u}^{\eps}(z-\eps^4,0) + \varphi(z-\eps^4)  + \ell^*_{\ep} - \eps^2v_z(z-\eps^4)  + \eps^2(1+\delta)v_z(z)+O(\eps^4)\\
& \geq  \delta\eps^2v_z(z)+ O(\eps^4)\\
& >  0,
\end{align*}
for all sufficiently small $\eps > 0$. Using $I^{\eps,\delta}(z^{\eps},0) = 0$, we deduce that $I^{\eps,\delta}$ attains a local minimum at some point $(\tilde{z}^{\ep},\tilde{\xi}^{\ep})$ with $|z_0-\tilde{z}^{\ep}| < r$ and $\tilde{\xi}^{\ep}\in \scr{G}(\tilde{z}^{\ep})$ for all  $\eps > 0$ sufficiently small. 

\textit{Step 3}: Now, we derive some limiting identities. Since, according to the previous argument, $|\tilde{\xi}^{\ep}|$ is uniformly bounded in $\eps$, there is a convergent subsequence $(\tilde{z}^{\ep_n}, \tilde{\xi}^{\ep_n})\to (\hat{z},\hat{\xi}),$ where $\hat{z} > 0$ and $\hat{\xi}\in \mathbb{R}$. Then,
\[
0  \geq \liminf_{n\to \infty} \eps_n^{-2}I^{\eps_n,\delta}(\tilde{z}^{\ep_n}, \tilde{\xi}^{\ep_n}) =  -\limsup_{n\to\infty} \overline{u}^{\eps}(\tilde{z}^{\ep_n}, \tilde{\xi}^{\ep_n}) + \varphi(\hat{z}) + C(\hat{z}-z_0)^4\]
by construction. Moreover:
\[-\limsup_{n\to\infty}\overline{u}^{\eps}(\tilde{z}^{\ep_n}, \tilde{\xi}^{\ep_n}) + \varphi(\hat{z}) + C(\hat{z}-z_0)^4 \geq  -u^*(\hat{z}) + \varphi(\hat{z}) + C(\hat{z} - z_0)^4 \geq 0.\]
So in fact, the inequalities must all be equalities. The strict maximality property of $u^*-\varphi$ at $z_0$ in turn gives $\hat{z} = z_0$ and $\hat{\xi}\in \overline{ \scr{G}(z_0)}$. Having chosen a particular subsequence, we may, without loss of generality, write $\eps$ instead of $\eps_n$. Using that $\underline{v}^{\eps}$ is a super-solution of the Dynamic Programming Equation \eqref{eq:dpe}, one obtains
\begin{align*}
0 & \leq \eps^{-2}(\beta \underline{v}^{\eps} - \scr{L}\psi^{\eps,\delta} - \tilde{U}(\psi^{\eps,\delta}_x))(\tilde{z}^{\eps},\tilde{\xi}^{\eps} )\\
&\leq \eps^{-2}(\beta \psi^{\eps,\delta} - \scr{L}\psi^{\eps,\delta} - \tilde{U}(\psi^{\eps,\delta}_x))(\tilde{z}^{\eps},\tilde{\xi}^{\eps})\\
& =  \underbrace{\frac{(\beta v - \scr{L} v - \tilde{U}(v_x))(\tilde{z}^{\eps},\tilde{\xi}^{\eps})}{\eps^{2}}}_{=:I_1} + \underbrace{\frac{\tilde{U}(v_x) - \tilde{U}(\psi^{\eps,\delta}_x)  }{\eps^2}}_{=:I_2} \\
&  \qquad -\underbrace{(\beta - \scr{L}) (\varphi(\tilde{z}^{\eps}) + \ell_{\ep}^* + C(\tilde{z}^{\eps} - z^{\ep})^4)}_{=:I_3} -\underbrace{\eps^2(1+\delta)(\beta w(\tilde{z}^{\eps},\tilde{\xi}^{\eps}) - \scr{L}w(\tilde{z}^{\eps},\tilde{\xi}^{\eps}))}_{=:I_4}.
\end{align*}
As $\eps\to 0$, we have
\begin{align*}
I_1 & \to -\frac{1}{2}|\sigma^{\top}\hat{\xi}|^2v_{zz}(z_0),\\
I_2  &\to  \tilde{U}'(v_z(z_0))\varphi_z(z_0) = - c_mz_0\varphi_z(z_0),\\
I_3&\to \mathcal{A}\varphi(z_0) - c_mz_0\varphi_z(z_0),\\
I_4&\to -\frac{1}{2}(1+\delta)\mbox{Tr}\left[\alpha(z_0)\alpha(z_0)^{\top}w_{\xi\xi}(z_0,\hat{\xi})\right].
\end{align*}
\textit{Step 4}: Combining the above limits with $\hat{\xi}\in \overline{\scr{G}(z_0)}$ yields the inequality
\begin{eqnarray*}
0 & \leq & -\frac{1}{2}|\sigma^{\top}\hat{\xi}|^2v_{zz}(z_0) - \mathcal{A}\varphi(z_0) + \frac{1}{2}(1+\delta)\mbox{Tr}\left[\alpha(z_0)\alpha(z_0)^{\top}w_{\xi\xi}(z_0,\hat{\xi})\right]\\
& = & a(z_0) - \mathcal{A}\varphi(z_0) + \frac{\delta}{2}\mbox{Tr}\left[\alpha(z_0)\alpha(z_0)^{\top}w_{\xi\xi}(z_0,\hat{\xi})\right].
\end{eqnarray*}
Finally, letting $\delta \to 0$ and using the local boundedness of $w_{\xi\xi}$ (cf.\ Proposition \ref{p.strip_estimate}) produces the desired inequality:
\[\Ac\varphi(z_0)\leq a(z_0).\]
\qed

\subsection{Viscosity Super-Solution Property}
\label{s.supersolution}

\begin{mythm}\label{t.supersolution}
$u_*$ is a viscosity super-solution of the Second Corrector Equation \eqref{e.au}.
\end{mythm}
\proof
Let $(z_0, \varphi)\in (0,\infty)\times C^2(\mathbb{R})$ be such that
\[0=(u_* - \varphi)(z_0) < (u_* - \varphi)(z),\quad \forall z>0,\ z\neq z_0.\]
To prove the assertion, we have to show
 \[\mathcal{A}\varphi(z_0) \geq a(z_0).\]

\textit{Step 1}: As before, we start by constructing the test function. Recall that
 \[\underline{u}^{\eps}(z,\xi) = \frac{v(z) - \overline{v}^{\eps}(z,\xi)}{\eps^2},\]
where $\overline{v}^\eps$ denotes the upper-semicontinuous envelope of the transaction cost value function $v^\eps$. By definition of the relaxed semi-limit $u_*$ and Lemma \ref{l.relaxed}, there exists a sequence $(z^{\eps}, 0)$ on the Merton line so that
\[z^{\eps}\to z_0 \mbox{ and }\underline{u}^{\eps}(z^\eps,0)\to u_*(z_0), \mbox{ as }\eps\to 0.\] 
Set 
\[\ell_{\eps}^*:= \underline{u}^{\eps}(z^{\eps})-\varphi(z^{\eps})\to 0\] 
and 
\[(x^{\eps},y^{\eps}):=(z^{\eps}-\pi_m\cdot\1_d z^{\eps}, \pi_m z^{\eps})\to (x_0,y_0):= (z_0 - \pi_m\cdot\1_d z_0, \pi_m z_0).\]
We localize by choosing $r > 0$ such that $B_r:=B_{r}(x_0,y_0)$ does not intersect the line $z=0.$ Define 
\[{\psi}^{\eps, \delta}(z,\xi) = v(z)-\eps^2(\varphi(z) + \ell^*_{\eps} - C(z-z^{\eps})^4) - \eps^4(1-\delta)w(z,\xi),\] 
where $C>0$ is chosen so large that 
\[I^{\eps,\delta}(z,\xi) := \overline{v}^{\eps}(z,\xi) - {\psi}^{\eps,\delta}(z,\xi) < 0,\quad \mbox{where $|z-z_0| \geq r/2,$ $\eps\in (0,\eps_0].$}\] 
Then, for all $\eps\in (0,\eps_0],$ we have
\begin{equation}\label{eq:supreme}
\sup_{\stackrel{|z-z_0| < r/2}{\xi\in \R^d}}I^{\eps,\delta}(z,\xi) < \infty,
\end{equation}
and, by construction,
 \[I^{\eps,\delta}(z^{\eps},0) = 0.\]

\textit{Step 2}: A priori, there is no reason that the supremum in \eqref{eq:supreme} should be achieved at any particular point, let alone that a maximizing sequence should converge as we send $\eps$ to zero. As a way out, we perturb the original test function to complete the localization. To this end, fix $\eps,\delta$ and let $(z_n, \xi_n)$ be a maximizing sequence of $I^{\eps,\delta}$. (Keep in mind that this sequence depends on $\eps,\delta$.) Set 
\[\alpha_n:= \frac{3}{2}\left(\sup_{\stackrel{|z-z_0| < r/2}{\xi\in \R^d}}I^{\eps,\delta}(z,\xi) - I^{\eps,\delta}(z_n, \xi_n)\right)\] 
and
\[h_n(\xi) = h(\xi - {\xi}_n),\] 
where 
\[ h(\xi) =  \begin{cases}
                                              \exp\left(1-\frac{1}{1-|\xi|^2}\right), &\mbox{if } |\xi| < 1, \\
                                              0 & \mbox{if } |\xi| \geq 1.
                                            \end{cases}
\] 
Notice that $\alpha_n\to 0$ as $\eps \to 0$. The modified test function is taken to be
 \[{\psi}^{\eps,\delta,n}(z,\xi) = \psi^{\eps,\delta}(z,\xi) - \alpha_nh_n(\xi),\] 
 so that 
 \[I^{\eps,\delta,n}(z,\xi):= ({v}^{\eps,\delta} - {\psi}^{\eps,\delta,n})(z,\xi) = I^{\eps,\delta}(z,\xi) + \alpha_n h_n(\xi).\] 
 By construction, each $I^{\eps,\delta,n}$ has a maximizer, say $(\hat{z}^{\eps}_n,\hat{\xi}^{\eps}_n)\in[z_0-r,z_0+r]\times \R^d.$ Observe that the rate of decay of $\alpha_n$ with respect to $\eps$ can be taken to be as fast as we wish, simply by choosing $n$ large enough. We will find it convenient to take $\alpha_{n_{\eps}} \leq \frac{1}{n_{\eps}}\exp(-\eps^{-1}),$ which can always be accomplished by relabeling, if necessary.

For any selection $\eps\mapsto\hat z^{\eps}_{n_{\eps}}$, it turns out that $z_0$ is the unique subsequential limit of $(\hat z^{\eps}_{n_{\eps}})_{\eps}$ as $\eps\to 0$. Indeed, note that since $(\hat z^{\eps}_{n_{\eps}})_{\eps}\subset[z_0-r,z_0+r]$, it contains a convergent subsequence. If $\hat{z}$ is the limit of such a subsequence, then
\[0  \leq  \eps_k^{-2}I^{\eps_k,\delta,n_k}(z^{\eps_k}_{n_k}, \xi^{\eps_k}_{n_k})\] 
which implies that 
\[0 \leq -u_*(\hat{z}) + \varphi(\hat{z}) - C(\hat{z}-z_0)^4.\] 
By the strict minimality of $u_*-\varphi$ at $z_0,$ we must have $\hat{z} = z_0.$

\textit{Step 3}: Next, we show that any sequence of maximizers $(\hat{z}^{\eps_k}_{n_k}, \hat{\xi}^{\eps_k}_{n_k})$ of $I^{\eps_k,\delta,n_k}$ where $\eps_k\to 0$ as $k\to \infty$, satisfying $\alpha_{n_k} \leq \frac{1}{k}\exp(-\eps_k^{-1}),$ is asymptotically contained in the no-trade region, that is,
\[(\overline{v}^{\eps_k} - \mbox{\bf M}\overline{v}^{\eps_k})(\hat{z}^{\eps_k}_{n_k}, \hat{\xi}^{\eps_k}_{n_k}) > 0\]
for all sufficiently large $k$. To see this, suppose by way of contradiction that instead
\begin{equation}\label{eq:contra}
0  \geq  \overline{v}^{\eps_k}(\hat{z}^{\eps_k}_{n}, \hat{\xi}^{\eps_k}_{n}) - (\mbox{\bf M}\overline{v}^{\eps})(\hat{z}^{\eps_k}_{n}, \hat{\xi}^{\eps_k}_{n})
 =  \overline{v}^{\eps_k}(\hat{z}^{\eps_k}_{n}, \hat{\xi}^{\eps_k}_{n}) - \overline{v}^{\eps_k}(\hat{z}^{\eps_k}_{n}-\eps^4, \tilde{\xi}^{\eps_k}_{n}),
 \end{equation}
for some $\tilde{\xi}^{\eps_k}_{n}.$ Such a point exists by the upper-semicontinuity and the boundedness of $\overline{v}^{\eps}$ at fixed wealths. Using the fact that $(\hat{z}^{\eps_k}_{n}, \hat{\xi}^{\eps_k}_{n})$ is a maximizer of $\overline{v}^{\eps_k} - {\psi}^{\eps_k,\delta,n_k}$ on $[z_0-r,z_0+r]\times \R^d$, one deduces that
\begin{eqnarray*}
0 & \geq & {\psi}^{\eps_k,\delta,n_k}(\hat{z}^{\eps_k}_{n}, \hat{\xi}^{\eps_k}_{n}) - {\psi}^{\eps_k,\delta,n_k}(\hat{z}^{\eps_k}_{n}-\eps^4, \tilde{\xi}^{\eps_k}_{n})\\
& = & \eps_k^4v_z(\hat{z}^{\eps_k}_{n}) + O(\eps_k^6) - \eps_k^4(1-\delta)[w(\hat{z}^{\eps_k}_{n},\hat{\xi}^{\eps_k}_{n}) - w(\hat{z}^{\eps_k}_{n}-\eps_k^4,\tilde{\xi}^{\eps_k}_{n})] - \eps_k^2\alpha_{n_k}[h_{n_k}(\tilde{\xi}^{\eps_k}_{n}) - h_{n_k}(\hat{\xi}^{\eps_k}_{n})]\\
& \geq & \delta\eps^4_{k}v_z(\hat{z}^{\eps_k}_n) + O(\eps_k^6)-\frac{2}{k}\eps_k^2\exp(-\eps_k^{-1}) \\
& > & 0,
\end{eqnarray*}
for all sufficiently large $k$ because $\delta > 0$. This contradicts \eqref{eq:contra}.

By the sub-solution property of $\overline{v}^{\eps_k}$ at $(\hat{z}^{\eps_k}_{n_k}, \hat{\xi}^{\eps_k}_{n_k}),$ for which we now write $(\hat{z}^{\eps_k}, \hat{\xi}^{\eps_k})$, one obtains the differential inequality
\begin{eqnarray*}
0&\geq& \left(\beta \overline{v}^{\eps_k} - \scr{L}\psi^{\eps_k,\delta,n_k} - \tilde{U}(\psi_x^{\eps_k,\delta,n_k})\right)(\hat{z}^{\eps_k}, \hat{\xi}^{\eps_k})\\
& \geq & \left(\beta \psi^{\eps_k,\delta,n_k} - \scr{L}\psi^{\eps_k,\delta,n_k} - \tilde{U}(\psi_x^{\eps_k,\delta,n_k})\right)(\hat{z}^{\eps_k}, \hat{\xi}^{\eps_k}).
\end{eqnarray*}

\textit{Step 4}:  We claim that $|\hat{\xi}^{\eps}|$ is uniformly bounded in $\eps\in (0,\eps_0]$. Expanding the above differential inequality into powers of $\eps_k$ leads to
\begin{eqnarray*}
0 & \geq & \eps_k^{-2}\left(\beta \psi^{\eps_k,\delta,n_k} - \scr{L}\psi^{\eps_k,\delta,n_k} - \tilde{U}(\psi_x^{\eps_k,\delta,n_k})\right)(\hat{z}^{\eps_k}, \hat{\xi}^{\eps_k})\\
& = & -\frac{1}{2}|\sigma^{\top}\hat{\xi}^{\eps_k}|^2v_{zz}(\hat{z}^{\eps_k}) - \alpha_{n_k}(\beta h_{n_k}(\hat{\xi}^{\eps_k}) - \scr{L}h_{n_k}(\hat{\xi}^{\eps_k}))\\
& & -(\beta(\varphi(\hat{z}^{\eps_k}) + \ell^*_{\eps_k} -C(\hat{z}^{\eps_k}-z^{\eps_k})^4)-\scr{L}(\varphi(\hat{z}^{\eps_k}) + \ell^*_{\eps_k} -C(\hat{z}^{\eps_k}-z^{\eps_k})^4))\\
& & -\eps_k^2(1-\delta)(\beta w(\hat{z}^{\eps_k},\hat{\xi}^{\eps_k}) - \scr{L}w(\hat{z}^{\eps_k},\hat{\xi}^{\eps_k})) - \frac{\tilde{U}(\psi^{\eps_k,\delta,n_k}_x) - \tilde{U}(v_x)}{\eps_k^2}.
\end{eqnarray*}
We proceed to estimate each term.  To this end, let $K = K(\beta,\mu,\sigma,r,\gamma)>0$ denote a sufficiently large generic constant. By Proposition \ref{p.strip_estimate}, we have
\begin{align*}
\eps_k^2(1-\delta)&(\beta w(\hat{z}^{\eps_k},\hat{\xi}^{\eps_k}) - \scr{L}w(\hat{z}^{\eps_k},\hat{\xi}^{\eps_k}))\\ &=-(1-\delta)\frac12\mbox{Tr}[\alpha(\hat{z}^{\eps_k})\alpha(\hat{z}^{\eps_k})^{\top}w_{\xi\xi}(\hat{z}^{\eps_k},\hat{\xi}^{\eps_k})] + (1-\delta)\eps \Rc_w(\hat{z}^{\eps_k},\hat{\xi}^{\eps_k})\\
& \leq  -(1-\delta)\frac12\mbox{Tr}[\alpha(\hat{z}^{\eps_k})\alpha(\hat{z}^{\eps_k})^{\top}w_{\xi\xi}(\hat{z}^{\eps_k},\hat{\xi}^{\eps_k})] + K(1+\eps_k |\hat{\xi}^{\eps_k}| +\eps_k^2 |\hat{\xi}^{\eps_k}|^2)
\end{align*}
and
\b*
\alpha_{n_k}|\beta h_{n_k}(\hat{\xi}^{\eps_k}) - \scr{L}h_{n_k}(\hat{\xi}^{\eps_k})| & \leq & \frac{K}{k}\exp(-\eps_{k}^{-1})(\eps_k^{-2}+\eps_k^{-1}|\hat{\xi}^{\eps_k}|+|\hat{\xi}^{\eps_k}|^2)
\e*
as well as
\[|(\beta - \scr{L}) (\varphi(\hat{z}^{\eps_k}) + \ell^*_{\eps_k} - C(\hat{z}^{\eps_k}-z^{\eps})^4) |  \leq  (1+\eps_k |\hat{\xi}^{\eps_k}| +\eps_k^2 |\hat{\xi}^{\eps_k}|^2).\]
Finally, 
\begin{equation*}
\eps_k^{-2}|\tilde{U}(\psi^{\eps_k,\delta,n_k}_x) - \tilde{U}(v_x)|\leq K.
\end{equation*}
We therefore conclude that
\begin{equation*}
0\geq -\frac{1}{2}|\sigma^{\top}\hat{\xi}^{\eps_k}|^2v_{zz}(\hat{z}^{\eps_k}) - K(1+\eps_k|\hat{\xi}^{\eps_k}|+\eps_k^2|\hat{\xi}^{\eps_k}|^2).
\end{equation*}
Recalling that $v_{zz} < 0$, it follows that the dominant term $-\frac{1}{2}|\sigma^{\top}\hat{\xi}^{\eps_k}|^2v_{zz}(\hat{z}^{\eps_k})$ is non-negative and therefore $|\hat{\xi}^{\eps_k}|$ must be uniformly bounded in $\eps_k$. Hence, along some subsequence we have $\hat{\xi}^{\eps_k}\to \hat{\xi}$ and $\hat{z}^{\eps_k}\to z_0.$ Sending $\eps_k \to 0$ gives
\begin{align*}
0 & \geq  -\frac{1}{2}|\sigma^{\top}\hat{\xi}|^2v_{zz}(z_0) - \mathcal{A}\varphi(z_0) + \frac{1}{2}(1-\delta)\mbox{Tr}\left[\alpha(z_0)\alpha(z_0)^{\top}w_{\xi\xi}(z_0,\hat{\xi})\right]\\
& =  a(z_0) - \mathcal{A}\varphi(z_0) - \frac{\delta}{2}\mbox{Tr}\left[\alpha(z_0)\alpha(z_0)^{\top}w_{\xi\xi}(z_0,\hat{\xi})\right].
\end{align*}
Finally, let $\delta\to 0$. Together with the $C^2$-estimates on $w$ (cf.\ Proposition \ref{p.strip_estimate}), it follows that the trace term disappears from the inequality. This yields
\[\mathcal{A}\varphi(z_0) \geq a(z_0),\]
thereby completing the proof.\qed

\subsection{Comparison for the Second Corrector Equation}\label{s.comparison}
A straightforward computation shows that $\Ac z^p = \nu_p z^p$ for some constant $\nu_p\in \R.$  If the Merton value function is finite, i.e.\ $c_m>0$, one readily verifies that $\nu_{1/2-\gamma} > 0$. Moreover, since the matrix $\alpha$ from \eqref{eq:posdef} is assumed to be invertible, the diffusion coefficient $A$ in \eqref{e.homothetic_corrector} is positive definite, so that $a(z) = a_0z^{1/2-\gamma}= [A:M]z^{1/2-\gamma} > 0.$ As a consequence, 
$$u(z)z^{-1/2+\gamma} = u_0 = \frac{a_0}{\nu_{1/2-\gamma}} > 0.$$ 

\begin{myrek}\rm{
Similarly, if $|\delta|\ll 1,$ then
\[\Ac (z^\delta u(z)) > 0,\quad \forall z>0.\] 
This observation is used in the proof of the Comparison Theorem \ref{t.comparison}. 
}\qed
\end{myrek}

In view of the explicit locally uniform upper bound \eqref{eq:upperbound} for $u^{\eps}$ from Theorem \ref{t.bounds}, the relaxed semi-limits $u^* ,u_*$ satisfy the growth constraint
\be
\label{e.growth}
0 \leq u^*(z), u_*(z)\leq C|z|^{1/2-\gamma}.
\ee
We therefore prove that the second corrector equation satisfies a comparison theorem in the class of non-negative functions satisfying this growth condition:

\begin{mythm}\label{t.comparison}
Let $v_1,v_2:(0,\infty)\to \mathbb{R}$ be positive and satisfy the growth constraint \eqref{e.growth}. If 
\[\Ac v_1 \leq a \leq \Ac v_2\] 
is satisfied in the viscosity sense, then \[v_1\leq u\leq v_2,\] where $u = u_0z^{1/2-\gamma}$. 
\end{mythm}

\proof
We just prove that the sub-solution is dominated by $u$; the second part of the assertion follows along the same lines. Let $v_1$ be a sub-solution to $\Ac v_1 \leq a$ satisfying the growth condition (\ref{e.growth}). We need to distinguish two cases:

\textit{Case 1}: Suppose $\gamma\neq 1/2.$ Set 
\[I(z) := v_1(z) - \phi_{\eta,\delta}(z),\] 
where 
\[\phi_{\delta,\eta}(z):= \delta u(z)^{1+\delta} + (1+\eta)u(z).\]
Then, for all sufficiently small $\delta,\eta > 0,$ we have $I(z) \leq 0,$ for all $z > 0.$ To see this, suppose on the contrary that, at some point, $I> 0$. Then, due to the growth restriction on $v_1$, $I$ will have a global maximum at some $\hat z \in \R_+$ and
 \[\Ac \phi_{\eta, \delta}(\hat z) \leq a(\hat z).\] 
However, by construction,
 \[\Ac \phi_{\eta, \delta}(\hat z) > a(\hat z).\] 
 Therefore, $I\leq 0$ everywhere. For $\delta,\eta \to 0$, this gives 
\[v_1(z)\leq u(z), \quad \forall z\in \R_+.\]

\textit{Case 2}: Now suppose $\gamma =1/2.$ Set instead 
\[\phi_{\delta,\eta}(z):= \delta z^{-\delta} + (1+\eta)u(z).\] 
The proof then follows along the same lines as in the first case.
\qed

\subsection{Proof of the Main Results}\label{s.mainproof}

We now conclude by proving the main results of the paper. 

\subsubsection*{Expansion of the Value Function}

\no{\bf{Proof of Theorem \ref{t.main}}.} We have shown in Theorem \ref{t.upperbound} that the relaxed semi-limits  $u^*$ and $u_*$ of \eqref{e.relaxed} exist, are functions of wealth only by Lemma \ref{l.relaxed} and, by Theorem \ref{t.bounds}, satisfy the growth condition \[0\leq u^*(z),u_*(z)\leq C|z|^{1/2-\gamma}.\] 
In view of Theorems \ref{t.subsolution} and \ref{t.supersolution}, 
\[\Ac u^* \leq a\leq \Ac u_*\] 
holds in the viscosity sense. As a result, Theorem \ref{t.comparison} gives 
\[u^*\leq u_*.\] 
The opposite inequality evidently holds by definition, therefore 
\begin{align}\label{e.equality_semilimits}
u_* = u^* = u.
\end{align}
The locally uniform convergence claimed in Theorem \ref{t.main} then follows directly from \eqref{e.equality_semilimits} and the definitions of $u_*, u^*.$
\qed

\subsubsection*{Almost Optimal Policy}\label{s.almost}

With the asymptotic expansion from Theorem \ref{t.main} at hand, we can now show that the policy from Theorem \ref{thm:strategy} is almost optimal for small costs. To this end, fix an initial allocation $(x,y)\in K_{\eps}$ and a threshold $0<\delta  < x+y \cdot \1_d$.  Consider the policy $\nu^{\eps}=(c^{\eps},\tau^{\eps}, m^{\eps})\in \Theta_{\eps}(x,y)$ from Theorem~\ref{thm:strategy}. If wealth falls below the threshold, another strategy is pursued (see Remark \ref{r.threshold}). More precisely, we choose controls $\nu^* = (c^*,\tau^*,m^*)\in \Theta_{\eps}(X_{\theta}, Y_{\theta})$ such that i) $\nu^{\eps}1_{[0,\theta)}+\nu^*1_{[\theta, 0)}\in \Theta_{\eps}(x,y),$ where $\theta$ is the first time the wealth process $Z^\eps=X^\eps+Y^\eps \cdot \1_d$ falls below level $\delta$, and ii) $\nu^*$ is $o(\eps^2)$-optimal on $[\theta,\infty)$ for each realization of $(X_\theta,Y_\theta)$. The main technical concern is whether this can be done measurably, but this will follow from a construction similar to the one performed in the proof of the weak dynamic programming principle \eqref{eq:dpp super}.  

Let $J^{\eps,\delta}(x,y)$ be the corresponding expected discounted utility from consumption:
\[J^{\eps,\delta}(x,y):=J(\nu) = \E\left[\int_0^\theta e^{-\beta s} U_{\gamma}(c_m Z^{\eps}_s)ds + e^{-\beta \theta}\int_{0}^{\infty}e^{-\beta t}U_{\gamma}(c^*_s)ds \right] \]
Then we have:

\begin{mythm}
There exists $\eps_{\delta} > 0$ such that, for all $0 < \eps \leq \eps_{\delta}$:
 \[J^{\eps,\delta}(x,y) \geq v(z)-\eps^2u(z)+ o(\eps^2),\quad \forall z=x+y\cdot\1_d \geq \delta.\]
 That is, the policy from Theorem \ref{thm:strategy} is optimal at the leading order $\eps^2$. 
\end{mythm}

\proof \textit{Step 1}: Set $$V^{\eps}(z,\xi) = v(z)-\eps^2u(z) - \eps^4(1+C\eps^2)w(z,\xi)$$ 
for some sufficiently large $C > 0$ to be chosen later. It\^o's formula yields
\begin{align*}
e^{-\beta \theta}V^{\eps}(X_{\theta},Y_{\theta}) =& V^{\eps}(x,y) + \int_0^{\theta}-e^{-\beta s}(\beta V^{\eps}(X_s,Y_s) - \scr{L}V^{\eps}(X_s,Y_s))ds\\
  &+\int_0^{\theta}e^{-\beta t}\mathbf{D}_y V^\eps(X_s,Y_s)^\top \sigma dW_s+ \sum_{t \leq \theta}(V^{\eps}(Z_{t},\xi_{t}) - V^{\eps}(Z_{t-},\xi_{t-})).
\end{align*}

\textit{Step 2}: We show that there are a sufficiently large $C>0$ and a sufficiently small $\eps_{\delta} > 0$ such that, for all $\eps\leq\eps_{\delta}$,
 \[\sum_{t \leq \theta}(V^{\eps}(Z_{t},\xi_{t}) - V^{\eps}(Z_{t-},\xi_{t-}))\geq 0.\] 
Expanding a typical summand, where $z \geq \delta$ and $\hat\xi\in \partial\scr{G}(z)$, we find that
\begin{align*}
V^{\eps}(z-\eps^4, 0) &-  V^{\eps}(z,\hat \xi)\\  
&=  v(z-\eps^4)-v(z) -\eps^2(u(z-\eps^4)-u(z)) -\eps^4(1+C\eps^2)(w(z-\eps^4,0) - w(z,\hat \xi))\\
& = -\eps^4v_z(z) -\eps^8 v_{zz}(\tilde{z_1}) + \eps^6u'(z) + \eps^{10} u''(\tilde{z_2}) + \eps^4v_z(z) + C\eps^6 v_z(z)\\
& \geq  C\eps^6 v_z(z) - \eps^8 v_{zz}(\tilde{z_1}) + \eps^6 u'(z) + \eps^{10}u''(\tilde{z_2})\\
& \geq  C\eps^6 v_z(z) - K\eps^6(z^{-1/2-\gamma} + \eps^2 \tilde{z_1}^{-1-\gamma} - \eps^4\tilde{z_2}^{-3/2-\gamma})\\
& >  0
\end{align*}
can be achieved for sufficiently small $\eps > 0$, uniformly in $z\geq \delta$, provided that $C$ is chosen large enough. (Here, the points $\tilde{z_1}, \tilde{z_2}\in [z-\eps^4,z]$ come from the Taylor remainders of $v$ and $u$, respectively.)

\textit{Step 3}: Next, establish that, for a suitable $k^*>0$, we have
\[\beta V^{\eps}(z,\xi) - \scr{L} V^{\eps}(z,\xi) \leq U(c_m z)(1+\eps^3k^*)\] for all $\eps < \eps_{\delta}$ and for all $z\geq \delta.$ 
Expanding the elliptic operator applied to $V^{\eps}$, we obtain
\b*
\beta V^{\eps} (z,\xi) - \scr{L} V^{\eps}(z,\xi) &\leq& -\frac12\eps^2 |\sigma^{\top}\xi|^2v_{zz}(z) + U(c_m z) - \eps^2\Ac u(z) + \frac12 \eps^2|\Rc_u(z,\xi)|\\
& & + (1+C\eps^2)\left(\eps^2\frac12\mbox{Tr}[\alpha(z)\alpha(z)^{\top}w_{\xi\xi}(z,\xi)] + \eps^3 |\Rc_w(z,\xi)|\right)\\
&& \\
& \leq & (1+\eps^3k^*)U(c_m z),
\e*
for sufficiently large $|k^*|$, where $k^*$ is positive for $\gamma < 1$ and negative for $\gamma>1$, thanks to the pointwise estimates on the remainder terms (cf.\ Remark \ref{r.estimates}) and the fact that $U(c_mz)$ is proportional to $z^{1-\gamma}.$ The argument for logarithmic utility is similar. The inequality therefore holds for all sufficiently small $\eps$ and for all $z\geq \delta.$

\textit{Step 4}: We now choose an appropriate control to use after time $\theta.$ Define the set 
\[
N_{\eps} = \{(x',y')\in K_{\eps}:\ \delta - 2\eps^4 \leq x'+y'\cdot \1_d\leq \delta+\eps^4 \}
\]
and, for each point $(x',y')\in N_{\eps}$, choose $(c,\tau,m)=\nu^{(x',y')}\in \Theta_{\eps}(x',y')$ such that 
\[
J(\nu^{(x',y')}) := \E\left[\int_0^{\infty}e^{-\beta s}U(c_t)dt \right] \geq v^{\eps}(x',y') - \eps^3.
\]
 We also define, for each $(x',y')\in N_{\eps}$, the set 
 \[
 R(x',y') = \{(\tilde x, \tilde y)\in N_{\eps}:\ \tilde x > x', \tilde y > y',\ \mbox{ such that }V^{\eps}(\tilde x, \tilde y) < V^{\eps}(x',y') + \eps^3 \}.
 \] 
 By construction we have 
 \[
 \bigcup_{(x',y')\in N_{\eps}}R(x',y')\supset NT^{\eps}\cap \{(\tilde x,\tilde y)\in K_{\eps}:\ \delta - \eps^4 \leq \tilde x + \tilde y\cdot \1_d\leq \delta  \}
 \] 
 and by compactness there is a finite sub-cover, say $(R(\zeta_n))_{n=1}^N$ for some $\zeta_1,\ldots, \zeta_N\in N_{\eps}$.

Now, define a mapping $\Ic: N_{\eps} \to \{1,\ldots, N\}$ which assigns to each point one of the neighborhoods in the subcover to which it belongs:
$$
\Ic(x^\prime,y^\prime):= \min\{n: (x^\prime,y^\prime) \in R(\zeta_n)\},
$$
and set
$$
\zeta(x^\prime,y^\prime):= \zeta_{\Ic(x^\prime,y^\prime)}.
$$
By the monotonicity of the value function,
\begin{equation}
v^\eps(\zeta) \le v^\eps(x^\prime,y^\prime), \quad \forall \ (x^\prime,y^\prime) \in R(\zeta).
\end{equation}
Also, since $V^{\eps}$ is smooth, each $R(\zeta)$ is open. Finally, define the following control $\nu^*_t\in \Theta_{\eps}(x,y)$:
$$
\nu^*_t:=
\left\{
\begin{array}{ll}
\nu^{\eps}_t, &\ {\mbox{if}}\ t \in [0,\theta],\\
\nu^\Nc_{t+\theta},&\ {\mbox{if}}\ t >\theta,\ {\mbox{with}}\
\Nc= \Ic((X,Y)_\theta).
\end{array}
\right.
$$ 
\
\
\textit{Step 5}: Piecing together the above estimates, and proceeding as in the proof of Theorem \ref{t.bounds} to get rid of the local martingale term, we obtain
\b*
V^{\eps}(x,y) & \leq &\E\left[e^{-\beta \theta}V^{\eps}(X_{\theta},Y_{\theta}) + \int_0^{\theta}e^{-\beta s}(\beta V^{\eps}(X_s,Y_s) - \scr{L}V^{\eps}(X_s,Y_s))ds\right]\\
& \leq & \E\left[e^{-\beta \theta}V^{\eps}(\zeta((X,Y)_\theta)) + \int_0^{\theta}e^{-\beta s}(1+\eps^3k^*)U(c_m Z_s^{\eps})ds\right]+\eps^3 \\
& \leq & \E\left[e^{-\beta \theta}J(\nu^{\Ic((X,Y)_\theta)}) + \int_0^{\theta}e^{-\beta s}(1+\eps^3k^*)U(c_m Z_s^{\eps})ds\right]+M_{\eps}+2\eps^3\\
& \leq & J^{\eps,\delta}(x,y) +M_{\eps} + \eps^3k^*v(z) + 2\eps^3,
\e*
where we have used in the last step that $k^* U$ is positive for $\gamma \neq 1$,\footnote{For logarithmic utility ($\gamma=1$), this follows similarly by additionally exploiting the estimate \eqref{eq:boundlog}.} and where
$$M_{\eps}:= \sup_{\stackrel{\delta-\eps^4\leq x'+y'\leq \delta}{(x',y')\in NT^{\eps}}}|V^{\eps}(x',y') - v^{\eps}(x',y')|.$$ 
The convergence results from Theorem \ref{t.main} imply that $M_{\eps}/\eps^2\to 0$ as $\eps\to 0.$  Since
\begin{align*}
J^{\eps,\delta}&\geq V^{\eps}(x,y) - M_{\eps} - \eps^3(2+k^*v(z))= v^{\eps}(x,y)-o(\eps^2),
\end{align*}
it follows that the proposed policy is indeed optimal at the leading order $\eps^2$.
\qed

\begin{appendix}

\section*{Appendix} 
\section{Pointwise Estimates}\label{app:A}

\begin{myprop}\label{prop:est}
There exists $K = K(\beta, \mu,r,\sigma,\gamma) > 0$ such that, for $k=0,1,2$ and $j=0,1,2$, we have
\[|\mbox{D}^k_{\xi}\partial^j_z w(z,\xi)|\leq K z^{-j-3k/4-\gamma}, \quad \mbox{for all $(z,\xi) \leftrightarrow (x,y)\in \mathrm{NT}^{\eps}$}.\] 
\end{myprop}

\begin{myrek}\label{r.estimates}\rm{
Proposition \ref{prop:est} yields the expansion
\[\eps^4(\beta w(z,\xi) - \scr{L}w(z,\xi)) = -\eps^2\frac12\mbox{Tr}[\alpha(z)\alpha(z)^{\top}w_{\xi\xi}(z,\xi)] + \eps^3 \Rc_w(z,\xi),\]
where the remainder term satisfies the bound
\[|\Rc_w(z,\xi)|\leq K (z^{1/4-\gamma}+\eps z^{-\gamma} + \eps^2z^{-1/4-\gamma}+\eps^3z^{-1/2-\gamma}).\]
In particular, \[\eps^4|\beta w(z,\xi) - \scr{L}w(z,\xi)| \leq  K(\eps^2z^{1/2-\gamma} + \eps^3z^{1/4-\gamma} + \eps^4z^{-\gamma} + \eps^5z^{-1/4-\gamma} + \eps^6z^{-1/2-\gamma}).\]
We can also expand
\[\eps^2(\beta u(z) - \scr{L} u(z)) = \eps^2(a(z) - c_mzu'(z)) + \eps^2\Rc_u(z,\xi),\]
with the following bound on the remainder:
\[|\Rc_u(z,\xi)|\leq K(\eps z^{1/4-\gamma} + \eps^2 z^{-\gamma}).\]
}
\end{myrek}

\proof This follows from tedious but straightforward computations, since all the functions and domains involved are known explicitly (compare \cite[Section 4.2]{soner.touzi.12} for a similar calculation).
\qed

\begin{myprop}\label{p.strip_estimate}
Let $\eps >0$ be given and consider $S:=[z_0-r_0,z_0+r_0]\times \mathbb{R}^d\subset K_{\eps}$, for some $z_0> r_0> 0.$  Then, given any $\Psi\in C^{2}(S)$ for which each restriction $D_{\xi}\Psi(z,\cdot)$ has compact support, there exists a $K>0,$ independent of $\eps,$ so that
\begin{align}\label{e.c2_finiteness}
\|\Psi\|_{C^2(S)} &\leq K
\end{align}
and
\begin{align}\label{e.elliptic_bound}
\eps^2|\beta \Psi(z,\xi) - \scr{L}\Psi(z,\xi)|& \leq  K(1+\eps|\xi|+\eps^2|\xi|^2),\quad \forall (z,\xi)\in S.
\end{align}
\end{myprop}

\proof This again follows from a tedious but straightforward calculation. \qed

\section{Proof of Theorem \ref{thm:visc}}\label{app:B}
\label{ap:dpp}

In this section we prove that for each fixed $\lambda,$ the value function $v^\lambda$ is a viscosity solution of the corresponding Dynamic Programming Equation~\eqref{eq:dpe} on the domain
$$\Oc_\lambda= \{ (x,y) \in K_\lambda\ :\
x+ y \cdot \1_d > 2\lambda\}.
$$
 As observed by Bouchard and Touzi \cite{bt.11}, a ``weak version'' of the dynamic programming principle is sufficient to derive the viscosity property via standard arguments (see for instance Chapter 7, in particular Theorem VII.7.1, in \cite{fleming.soner.06}). Rather than checking the abstract hypotheses of \cite{bt.11}, we present, for the convenience of the reader, a direct proof of the weak dynamic programming principle in our specific setting, using the techniques of \cite{bt.11}.

To this end, fix $(x,y) \in \Oc_\lambda$, $\delta >0$, let $B_{\delta}(x,y)\subset \R^{d+1}$ denote the ball of radius $\delta$ centered at $(x,y)$, and set
$$
K(x,y;\delta)_\lambda:=
\{ (x^\prime,y^\prime)\ :\ 
x+ y \cdot \1_d -\delta-\lambda \le
x^\prime+ y^\prime \cdot \1_d \le
x+ y \cdot \1_d +\delta\}.
$$
Take $\delta > 0$ sufficiently small so that
$K(x,y;2\delta)_\lambda \subset \Oc_\lambda$. For any investment-consumption policy $\nu$, define $\theta$ as the exit time of the corresponding state process $(X,Y)$  from $B_{\delta/2}(x,y)$. (Following standard convention, our notation does not explicitly show the dependence of $\theta$ on $\nu$.) It is then clear that
 $$
 (X,Y)_{\theta^-} \in \overline {B_{\delta/2}(x,y)},\quad
 {\mbox{and}} \quad
 (X,Y)_{\theta} \in K(x,y,\delta)_\lambda.
 $$
The following weak version of the DPP is introduced in \cite{bt.11}:

Let $\varphi$ be a smooth and bounded function on $ K(x,y,2\delta)_\lambda$, satisfying
$$
0=(v^\lambda-\varphi)(x,y)= \max\left\{(v^\lambda-\varphi)(x^\prime,y^\prime):\ 
(x^\prime,y^\prime) \in K(x,y,2\delta)_\lambda\right\}.
$$
Then, we have
\begin{equation}
\label{eq:dpp sub}
v^\lambda(x,y) \le \sup_{\nu \in \Theta_\lambda (x,y)}\E\left[
\int_0^\theta e^{-\beta t} U(c_t)dt + 
e^{-\beta \theta} \varphi( (X,Y)_{\theta}) \right].
\end{equation}
The restriction to bounded test functions $\varphi$ is possible since by \eqref{e.bounds_veps}, $v^{\lambda}$ is bounded on $K(x,y;2\delta)_{\lambda}$.

Conversely, let $\varphi$ be a smooth function 
bounded on $ K(x,y,2\delta)_\lambda$, satisfying
$$
0=(v^\lambda-\varphi)(x,y)= \min\left\{(v^\lambda-\varphi)(x^\prime,y^\prime):\ 
(x^\prime,y^\prime) \in K(x,y,2\delta)_\lambda\right\}.
$$
Then, we have
\begin{equation}
\label{eq:dpp super}
v^\lambda(x,y) \ge \sup_{\nu \in \Theta_\lambda (x,y)}\E\left[
\int_0^\theta e^{-\beta t} U(c_t)dt + 
e^{-\beta \theta} \varphi( (X,Y)_{\theta}) \right].
\end{equation}

\noindent
{\bf{Proof of the Weak Dynamic Programming Inequalities \reff{eq:dpp sub} and \reff{eq:dpp super}.}}
\vspace{5pt}

We start with the proof of \reff{eq:dpp sub}.
For any policy $\nu \in \Theta_\lambda(x,y)$,
let $\nu^\theta=(c^\theta,\tau^\theta,m^\theta)$ be its restriction to the (stochastic) interval $[\theta, \infty)$.
Then, $\nu^\theta \in \Theta_\lambda((X,Y)_{\theta})$, $P$-a.s. Therefore, 
\[\E \left[\int_0^{\infty}e^{-\beta t}U(c^{\theta}_t)dt \Big\vert \scr{F}_\theta \right] \leq v^\lambda((X, Y)_{\theta}) \leq \varphi((X, Y)_{\theta}),\quad P\mbox{-a.s.,}  \]
holds by definition of the value function and because $(X,Y)_\theta \in K(x,y,\delta)_\lambda$ lies in the set $K(x,y,2\delta)_\lambda$ where $\varphi$ dominates $v^\lambda$ by definition. As a result, for any $\nu=(c,\tau,m) \in \Theta_\lambda(x,y)$,
\begin{eqnarray*}
\E \left[\int_0^\infty e^{-\beta t} U(c_t)dt\right] 
&= & \E \left[\int_0^\theta e^{-\beta t} U(c_t)dt +
e^{-\beta \theta}\int_\theta^\infty e^{-\beta (t-\theta)} U(c_t)dt \right]\\
&\le & \E\left[\int_0^\theta e^{-\beta t} U(c_t)dt + e^{-\beta \theta}\E \left[\int_0^{\infty}e^{-\beta t}U(c^{\theta}_t)dt\Big\vert \Fc_{\theta}\right]  \right]\\
& \le & \E \left[\int_0^\theta e^{-\beta t} U(c_t)dt +
e^{-\beta \theta}\varphi((X,Y)_{\theta})\right].
\end{eqnarray*}
By taking the supremum over all policies $\nu$, we arrive at \reff{eq:dpp sub}.

To prove \reff{eq:dpp super}, set
$\V$ to be the right hand side of \reff{eq:dpp super}, that is:
$$
\V:=  \sup_{\nu \in \Theta_\lambda (x,y)}\E\left[
\int_0^\theta e^{-\beta t} U(c_t)dt + 
e^{-\beta \theta} \varphi( (X,Y)_{\theta}) \right].
$$
For any $\eta >0,$ we can choose $\nu^\eta \in \Theta_\lambda(x,y)$
satisfying
\begin{equation}
\label{eq:est0}
\V \le \eta+\E\left[
\int_0^\theta e^{-\beta t} U(c^\eta_t)dt + 
e^{-\beta \theta} \varphi( (X,Y)_{\theta}) \right].
\end{equation}
We have already argued that $(X,Y)_{\theta} \in  K(x,y,\delta)_\lambda$. Our next step is to construct a countable open cover of $K(x,y,\delta)_\lambda$.  For every point $\zeta = (\tilde x, \tilde y)\in K(x,y,2\delta)_\lambda$, set
$$
R(\zeta):= R_\eta(\tilde x, \tilde y)=\{ (x^\prime,y^\prime) \in K(x,y,2\delta)_\lambda\ :\
x^\prime >\tilde x, \ y^\prime > \tilde y, \
\varphi(x^\prime,y^\prime) < \varphi(\tilde x, \tilde y) + \eta\}.
$$
By the monotonicity of the value function,
\begin{equation}
\label{eq:est1}
v^\lambda(\zeta) \le v^\lambda(x^\prime,y^\prime), \quad \forall \ (x^\prime,y^\prime) \in R(\zeta).
\end{equation}
Also, since $\varphi$ is smooth, each $R(\zeta)$ is open and
$$
K(x,y,\delta)_\lambda \subset \bigcup_{\zeta \in K(x,y,2\delta)_\lambda} R(\zeta).
$$
Hence, by the Lindel\"of covering lemma, we can extract a countable subcover
$$
K(x,y,\delta)_\lambda \subset \bigcup_{n\in \N}\ R(\zeta_n).
$$
Now, define a mapping $\Ic: K(x,y,\delta)_\lambda \to \N$ which assigns to each point one of the neighborhoods in the subcover to which it belongs:
$$
\Ic(x^\prime,y^\prime):= \min\{n: (x^\prime,y^\prime) \in R(\zeta_n)\},
$$
and set
$$
\zeta(x^\prime,y^\prime):= \zeta_{\Ic(x^\prime,y^\prime)}.
$$
By definition, these constructions imply
\begin{equation}
\label{eq:est2}
\varphi(x^\prime,y^\prime) \le \varphi(\zeta(x^\prime,y^\prime)) + \eta, \quad
\forall \ (x^\prime,y^\prime) \in K(x,y,\delta)_\lambda.
\end{equation}
As a final step, for each positive integer $n$,
we choose a control $\nu^n \in \Theta_\lambda(\zeta_n)$ 
so that
\begin{equation}\label{eq:nun}
v^\lambda(\zeta_n) \le J(\nu^n) + \eta,\quad
{\mbox{where }}
J(\nu):=\E \left[
\int_0^\infty e^{-\beta t} U(c_t)dt \right] \ {\mbox{for any control $\nu=(c,\tau,m)$}}.
\end{equation}
By monotonicity, $\nu^n \in  \Theta_\lambda(x^\prime,y^\prime)$
for every $(x^\prime,y^\prime) \in R(\zeta_n)$.  We now define a composite strategy  $\nu^*$, which follows the policy $\eta$ satisfying \eqref{eq:est0} until the corresponding state process $(X,Y)$ leaves $B_{\delta/2}(x,y)$ at time $\theta$. It then switches to the policy $\nu^n$ corresponding to the index $n$ which the state process is assigned by the mapping $\mathcal{I}$:
$$
\nu^*_t:=
\left\{
\begin{array}{ll}
\nu^\eta_t, &\ {\mbox{if}}\ t \in [0,\theta],\\
\nu^\Nc_{t+\theta},&\ {\mbox{if}}\ t >\theta,\ {\mbox{with}}\
\Nc= \Ic((X,Y)_\theta).
\end{array}
\right.
$$ 
This construction ensures $\nu^* \in \Theta_\lambda(x,y)$. Hence, it follows from the definitions of the value function and $\nu^*$, \eqref{eq:nun} and $v^\lambda \ge \varphi$ (which holds for $(X,Y)_\theta \in K(x,y,2\delta)_\lambda$ by definition of $\varphi$), as well as \eqref{eq:est2} and \eqref{eq:est0} that
\begin{align*}
v^\lambda(x,y) \ge  J(\nu^*) &= \E\left[ \int_0^\theta e^{-\beta t} U(c^\eta_t)dt + 
 e^{-\beta \theta}\left. \E\left[\int_0^{\infty}e^{-\beta t}U(c_t^{\Nc,\theta})dt  \right| \Fc_{\theta}  \right]  \right]\\
 &\ge \E\left[ \int_0^\theta e^{-\beta t} U(c^\eta_t)dt + 
 e^{-\beta \theta} [\varphi(\zeta((X,Y)_\theta) - \eta]\right]\\
  &\ge \E\left[ \int_0^\theta e^{-\beta t} U(c^\eta_t)dt + 
 e^{-\beta \theta} [\varphi((X,Y)_\theta) -2 \eta]\right]\\
 &\ge  \V - 3 \eta.
 \end{align*}
 Since $\eta$ was arbitrary, this establishes \eqref{eq:dpp super} and thereby completes the proof.
 \qed

\end{appendix}

\bibliographystyle{abbrv}
\bibliography{tractrans}

\end{document}